\documentclass[trackchanges]{aastex63}

\usepackage{graphicx}
\usepackage{rotating}
\usepackage{hyphenat}
\usepackage{footnote}
\usepackage{url}
\usepackage{color,soul}
\usepackage{lineno}
\usepackage{amsmath}
\usepackage{tabularx}
\usepackage{comment}

\usepackage{hyperref}
\usepackage{amsmath}
\usepackage{MnSymbol}

\usepackage{float}
\usepackage{url}
\usepackage{CJK}

\def\apjs{{ApJS}}
\def\apjl{{ApJL}}

\def\aap{{A\&A}}

\newcounter{supp}[section]

\submitjournal{PSJ}
\accepted{December 15, 2025}

\shorttitle{Rocky planet EOS+mineral suites}

\graphicspath{{./}{}}

\begin{document}

\title{Toward Reliable Interpretations of Small Exoplanet Compositions: \\ Comparisons and Considerations of Equations of State and Materials Used in Common Rocky Planet Models}

\correspondingauthor{Joseph Schulze}
\email{jschulze@lsu.edu}

\author[0000-0003-3570-422X]{Joseph G. Schulze}
\affiliation{Department of Physics \& Astronomy, Louisiana State University, 202 Nicholson Hall
Baton Rouge, LA 70803, USA}

\author[0000-0003-0595-5132]{Natalie R. Hinkel}
\affiliation{Department of Physics \& Astronomy, Louisiana State University, 202 Nicholson Hall
Baton Rouge, LA 70803, USA}

\author[0000-0001-5753-2532]{Wendy R. Panero}
\affiliation{Division of Earth Sciences, National Science Foundation, 2415 Eisenhower Ave, Alexandria, VA, 22314, USA}

\author[0000-0001-8991-3110]{Cayman T. Unterborn}
\affiliation{Southwest Research Institute, 6220 Culebra Road, San Antonio, TX 78238, USA}

\begin{abstract}
The bulk compositions of small planets ($R_p< 2 \mathrm{R}_\oplus$) are directly linked to their formation histories, making reliable compositional constraints imperative for testing models of planet formation and evolution. Because exoplanet interiors cannot be directly observed, their make-up must be inferred from mass–radius–composition models that link assumed stellar abundances to the direct observables: planetary mass and radius. There are a variety of such models in the literature, each adopting different equations of state (EOS) to describe the materials' properties at depth and varying assumptions about the minerals present within the planets. These EOS+mineral suites provide the foundations for compositional inferences, but they have not yet been systematically compared. In this work, we review several suites, with a detailed description of the basic structure, mineral physics, and materials within standard small planet models. We show that EOS+mineral suites predict planet densities whose differences are comparable to current observational uncertainties, which present a challenge for robustly interpreting and classifying small planets. We apply a powerful small-planet characterization framework, which illustrates that variations among EOS+mineral suites lead to inconsistent conclusions for both individual planets and sample-level demographics. Our results demonstrate the need for more careful considerations of the materials and EOS used in mass-radius-composition models, especially given the current focus on finding and characterizing potentially habitable rocky planets. We conclude with recommendations for best practices so that future interpretations of small planets and their formation are accurate and consistent. 
 \end{abstract}


\section{Introduction}
\label{sect:intro}
Small planets ($R_p< 2 \mathrm{R}_\oplus$) outside the Solar System (or exoplanets) are some of the most common planets in the galaxy \citep[e.g.,][]{Fulton_2017, Neil_2020, cloutier_2020}. Drawing upon the Solar System's rocky planets and icy moons, small exoplanets are thought to be made primarily from the \textit{major rock-building elements}: Fe, Mg, and Si. Planets form from the same gas-dust disk as their host star(s). Leftover debris orbit around the new star, where the first materials to condense into solids from the disk are Ca- and Al-bearing minerals (e.g., corundum, Ca-silicate-perovskite, anorthite) followed by Fe-alloys and Mg-bearing silicates (e.g., forsterite, enstatite, and bridgmanite). In the inner regions of the gas-dust disk, where the temperatures are too high for H$_2$O to solidify, planet-building materials are dominated by Fe and silicates. As planetesimals accrete enough material to gravitationally reshape, they compositionally differentiate with Fe settling to the center to form a core surrounded by a silicate mantle. While Ca and Al are incorporated into the mantle, they are less abundant than Si and Mg by a factor of $\sim10-16$ within the disk \citep{Lodders03}, and are therefore referred to as minor mantle elements. Together with O, the major rock-building elements make up 95 mol \% of Earth \citep[][]{McDonough_2003}. As such, we expect planets forming in these inner, hot regions of the disk to be effectively volatile- and water-free rocky planets with Fe cores and silicate mantles. Some distance away from the host star, the temperatures of the gas-dust disk drop below the condensation temperature of H$_2$O, commonly referred to as the snow-, frost-, or ice-line. Beyond the snow-line, a disk of solar composition is expected to have approximately equal amounts of H$_2$O-ice and rocky materials by mass \citep[][]{Lodders03}, where small planets may have water mass fractions (WMF) of $\sim0.5$ (or 50\%), such as icy moons like Ganymede or Callisto \citep[e.g.,][]{Showman_1999, Kuskov_2005}.

With a structure of Fe-dominated core, silicate mantle, and outer hydrosphere, a small planet's bulk density is determined primarily by two compositional parameters: the WMF and molar Fe/Mg ratio. The WMF is the relative amount of rocky materials (core+mantle) to H$_2$O-ice and/or liquid water on the surface, with $\mathrm{WMF}=0$ corresponding to a H$_2$O-free rocky planet. Increasing the WMF at a fixed planet mass trades rocky material for lower-density H$_2$O, which results in a reduction in the average, or bulk, density of the planet. The Fe/Mg ratio is the primary control on the relative ratio of the denser Fe core to the less-dense silicate mantle. At a fixed planet mass, increasing the Fe/Mg ratio increases the mass ratio of core-to-mantle: the planet must trade less dense mantle material for more dense core material to maintain its mass leading to an increase in bulk planet density. 

The Fe-dominated cores of small planets may also be enriched in some amount of `light elements' (compared to the molar mass of Fe) like H, C, O, Si, and S, but are soluble in Fe during differentiation. While the exact elements remain unknown, Earth's core is estimated to have $\sim 10$ wt \% light elements and Mars' core, while less precisely known, is estimated to have $\sim 20$ wt \% \citep[e.g.,][]{McDonough_2003, hirose2021light,Irving_2023, samuel2023geophysical, bi2025seismic}. Introduction of light elements into the core lowers the density of the core relative to pure Fe which, in turn, reduces the bulk density of the planet. Similarly, imperfect settling of Fe into the core and oxidation of Fe during planetary formation retains oxidized Fe in the mantle at the expense of Fe in the core. It's estimated that the Earth has $\sim8$ wt \% FeO in its mantle \citep[][]{McDonough_2003} while Mars has $\sim18$ wt \% \citep[][]{Wanke94_Chem_and_Acc_Hist_Mars}. This results in competing effects: the relative mass of the core is reduced, implying a reduction in planet density, while the density of the mantle is increased, implying an increase in planet density.

The exact composition of a small planet is inextricably linked to its formation history. The major rock-building elements condense into solids over a narrow temperature range \citep[$\sim1310-1350$ K,][]{Lodders03}, meaning the bulk Fe/Mg ratios of rock-dominated planets should not be significantly altered relative to the Fe/Mg ratio of their host stars \citep[][]{thiabaud2015}. In other words, we expect a nearly 1:1 relationship between the relative molar ratios of moderately-refractory and refractory elements (e.g, Fe, Mg, Si, Ca, Al) in a star and the make-up of its rocky planet, known as the rock-star relationship. This relationship is true when considering the Earth and Mars\footnote{While the composition of Venus is poorly constrained, it is expected to be consistent with Earth and therefore the rock-star relationship \citep{Zharkov83_IntStruct_Venus}.} with respect to the Sun, and is generally true for small exoplanets with precise densities and measured host abundances of the major rock-building elements \citep[e.g.,][]{thiabaud2015, schulze_2021, Adibekyan_2021, Unterborn23, Adibekyan_2024}. The rock-star relationship does not hold for Mercury, which has a molar Fe/Mg ratio $\gtrsim 5\times$ greater than that of the Sun \citep[][]{Morgan1980, nittler, Hauck_2013, Lodders03}, meaning it is significantly Fe-enriched. There is also growing evidence for exoplanets more massive than Mercury with similar levels of Fe-enrichment (or super-Mercuries) like Kepler-107 c, HD 137496 b, and Kepler-406 b \citep[][]{schulze_2021, Adibekyan_2021, Silva_2022, Unterborn23, Adibekyan_2024}. Such Fe-enriched planets are hypothesized to form through large-scale mantle-stripping collisions \citep[e.g.,][]{Marcus10_MinRad_SEs} or processes that separate Fe from silicates within the disk \citep[e.g., nucleation of solid particles + streaming instability, photophoresis, magnetic boost:][respectively]{Johansen_2022, Wurm_2013, Kruss_2020}. In short, the rock-star relationship is the chemical basis on which we try to understand exoplanetary interiors, with deviations from it offering insights into physical processes not generally included in stand models of small planet formation \citep[][]{schulze_2021}.

There are currently $>1600$ \textit{confirmed} small exoplanets on the NASA Exoplanet Archive\footnote{\href{https://exoplanetarchive.ipac.caltech.edu/index.html}{https://exoplanetarchive.ipac.caltech.edu/index.html} as of 27 AUG 2025. A planet was identified as a `confirmed small planet' if its default NEA radius was $R_p<2R_\oplus$ and it was labeled as uncontroversial.}\citep[NEA,][]{NEA_reference}. Of these, 99\% have orbital periods $\leq100$ days and 85\% orbit nominally Sun-like, or FGK-type, stars\footnote{Defined here as having an effective temperature between 4000--6500 K, following the temperature range of the \textit{Gaia} FGK benchmark set \citep[][]{Heiter_2015}.}. While the location of the snow-line is not static and depends on the properties of the host star and disk, it is generally predicted to remain exterior to an Earth-like orbital distance (i.e., orbital period $>365$ days) around Sun-like stars \citep[e.g.,][]{Hayashi_1981, Podolak, Martin_2012, Bailli_2015, Zhang_2015}. If most of the confirmed small exoplanets formed in situ, they would have likely formed interior to their system's respective snow-line, and therefore we expect that their  WMF $\approx$ 0. When a small planet has a WMF $>$ 0, it suggests either migration from beyond the snow-line or late-stage volatile delivery.

\subsection{Inferring the compositions of small exoplanets}
Unfortunately, the composition of a small exoplanet is not \textit{directly} observable. Instead, it must be \textit{inferred} from the planet's measured mass and radius, and therefore density, through mass-radius-composition models of which there are many in the literature, such as: \citet{Seager_2007, Sotin_2007,  Valencia_2007, Zeng_Sasselov_2013, Dorn15, Zeng16_emperical_MR_function, dorn17_bayes, Magrathea_2022,  Unterborn23}. A mass-radius-composition model maps out planet radius as a function of mass, or vice versa, for a fixed input planet composition accounting for the self-compression of planetary materials with pressure. For the same planet composition, a 10$M_\oplus$ planet will naturally be more dense than a 1$M_\oplus$ planet because its higher gravity compresses the planetary materials more, leading to an overall increase in their average density. For example, on a mass-radius diagram, an Earth-like model will pass through/near the masses and radii of Earth, Mars, and Venus, but not Mercury. Generally, the problem is inverted to infer the composition of an exoplanet: the mass and radius are specified a priori due to detection methods and the compositional parameters are varied until a mass-radius-composition model is found that passes through the input mass and radius.

Overall, small exoplanets are interpreted as belonging to one of three broad classes based on their inferred Fe/Mg ratios and WMFs, with each class offering unique insights into planet formation: (1) Super-Mercuries, with high Fe/Mg ratios, can help constrain the frequency of giant impacts or their relative importance compared to compositional sorting within the disk; (2) Exoplanets that follow the rock-star relationship can help constrain the number of planets whose composition mirror that of their host and identify those most likely to be Earth analogs; (3) Water worlds and other surface-volatile-rich planets with orbital distances of $\lesssim1$ AU can provide constraints on planet migration rates from beyond the snow-line or late-stage volatile delivery. Accurate constraints on small exoplanet compositions are imperative to properly interpret and understand the processes that shape them.

The first step to interpreting a small planet with both a mass and radius measurement is to select an existing mass-radius-composition model or build a new one. Each model must first make different a priori assumptions about the minor aspects of the planet's composition, such as whether there are light elements in the core and/or Fe within the mantle (Sect. \ref{sect:intro}-\ref{sect:struct_and_minerology_overview}). Next, it must be decided what minerals/materials make up the planet model (Sect. \ref{sect:struct_and_minerology_overview}) which are often informed by the major rock-building host star abundances or Earth. For each selected planetary material, a material-specific equation of state (EOS) must be chosen that links its physical properties, such as density or volume, to the pressure and temperature within the planet (Sect. \ref{sect:common_eos}). For a given material, there are generally many EOS within the mineral physics literature to choose from \citep[e.g., for $\epsilon-$Fe, a high-pressure allotrope of solid Fe:][and others]{Mao_1990, Anderson_2001, Uchida_2001, belonoshko_2010, Bouchet_2013, Smith18_Fe_EoS}. Here we refer to the set of compositional assumptions, minerals/materials and EOS underlying a mass-radius-composition model as an EOS+mineral suite. Finally, the EOS+mineral suite must be applied to a planet-building algorithm that iteratively solves the structure equations that govern planetary interiors until a self-consistent solution is found (Sect. \ref{sect:planet_builder}). Together, differences between mass-radius-composition models lead to differences in the predicted densities of small planets which can lead to inconsistencies in their interpretations. 

In this work, we review the EOS+mineral suites adopted in 10 commonly used mass-radius-composition models for purely rocky planets with iron-dominated cores and silicate mantles (Sect. \ref{sect:selected_EOS_mineral_suites}). That is, we limit our scope to H$_2$O- and volatile-free rocky planets where Fe/Mg is the dominant composition parameter controlling planet bulk density. Further, most rocky planet EOS+mineral suites assume the mantles are entirely solid which is indeed the case for our selected suites. As such, the EOS+mineral suites in this work are applicable only to rocky planets with no (or negligible) atmospheres and solid, dry silicate mantles surrounding iron-dominated cores, which may not be the case for many known small planets \citep[e.g.,][]{Zilinskas_2022, Boley_2023, Young_2024}. We will address how suites for H$_2$O change the inferred WMF of water-worlds in future works.

Next, we quantify the differences in predicted material densities from the various EOS adopted for core materials and mantle minerals over a range of pressures expected within small, rocky planets (Sect. \ref{sect:direct_EOS_comparisons}), and how these differences influence the predicted bulk densities of rocky planets at different mass scales and bulk Fe/Mg (Sect. \ref{sect:effects of EOS+mineral suite on density}). We outline a powerful small planet characterization scheme that incorporates a large sample of stellar Fe/Mg to predict the likely compositional range of rocky planets allowing for statistically robust identification of compositional outliers (e.g., super-Mercuries or water-worlds; Sect. \ref{sect:planet_interpretations_and_inferences}). We then apply the scheme to a sample of small planets with precisely determined densities ($<25\%$ uncertainty) and find that different EOS+mineral suites do indeed give inconsistent interpretations for a large number of planets (Sect. \ref{sect:planet_interpretation_results}): for example, a given planet is interpreted as a super-Mercury or Earth-like depending on the adopted suite.

\section{Basic Structure and Mineralogy of Rocky Planets}
\label{sect:struct_and_minerology_overview}

\begin{figure}[h]
    \centering
    \includegraphics[width=0.85\linewidth]{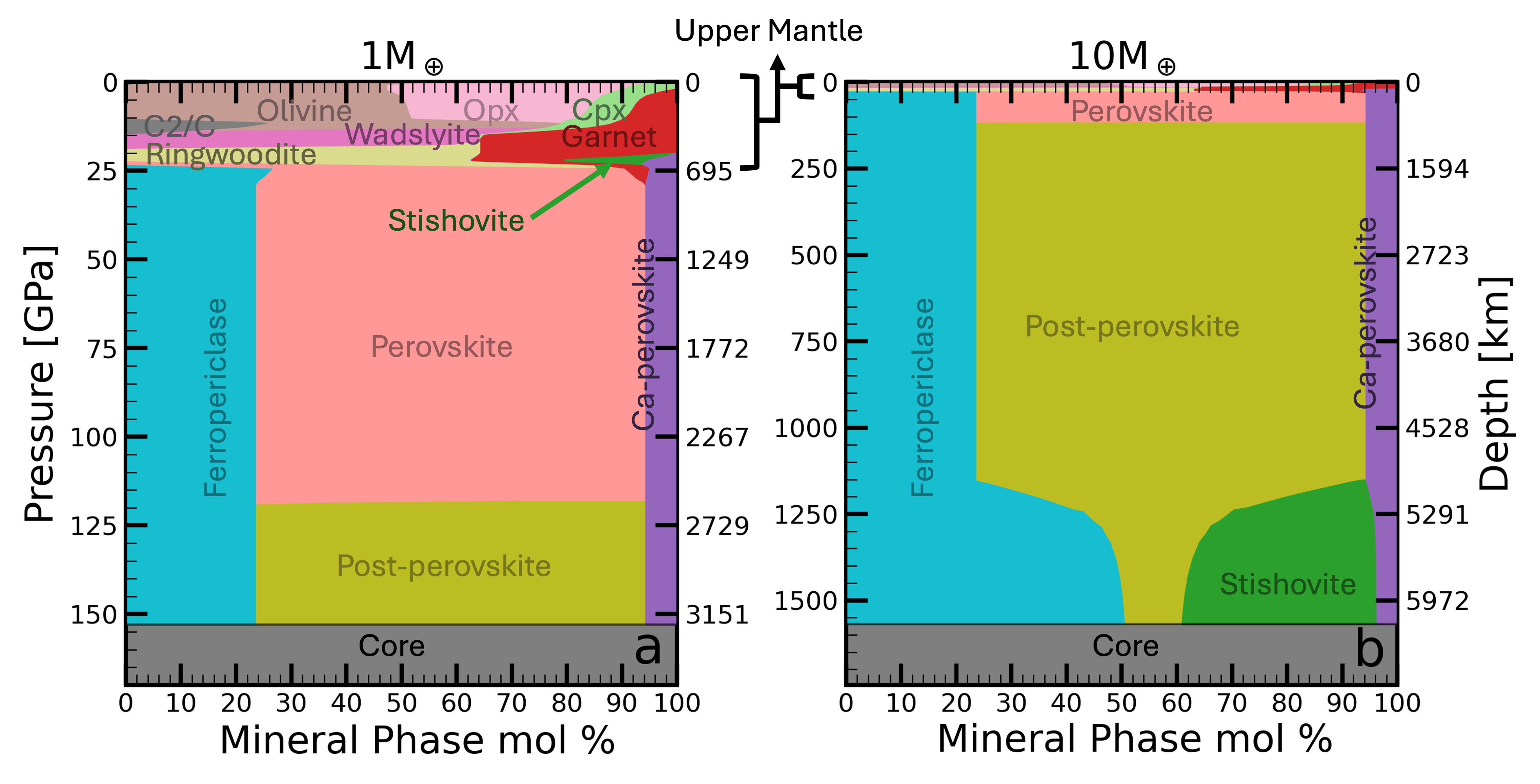}
    \caption{Representative sections of (a) 1$\mathrm{M_\oplus}$ and (b) 10$\mathrm{M_\oplus}$ planets as a function of depth and pressure. Both were created using the \texttt{Perple\_X} software with the thermodynamic database of \citet{Stixrude_2011} (see e.g., Sect. \ref{subsection:exoplex}) assuming an Earth-like composition (Fe/Mg = 0.9, Si/Mg = 0.9, Ca/Mg = 0.07, Al/Mg = 0.09, and 8 wt \% mantle FeO). Cpx = clinopyroxene, Opx = orthopyroxene, and C2/C = C2/C orthopyroxene. The garnet is majoritic garnet. Perovskite is silicate-perovskite.}
    \label{fig:mantle_section}
\end{figure}

As a mineral is compressed and/or heated, it may rearrange into a more energetically favorable, or stable, crystal structure becoming a different mineral in the process. These mineral restructurings are phase transitions which act to alter the material's properties, e.g., increase its density and decrease its compressibility. Different minerals with the same chemical formulas but different crystal structures and material densities are polymorphs. In Fig. \ref{fig:mantle_section}, we show mineral phase sections for a 1$M_\oplus$ (Fig. \ref{fig:mantle_section}a) and 10$M_\oplus$ (Fig. \ref{fig:mantle_section}b) planet with compositions similar to Earth.

The uppermost region of a rocky planet's mantle, or the upper mantle (shown in brackets in Fig. \ref{fig:mantle_section}), is limited to pressures $\lesssim25$ GPa and characterized by a series of rapid phase transitions. At pressures $\lesssim14$ GPa, the upper mantle is volumetrically dominated by olivine and orthopyroxene (labeled as `Opx'). Olivine has the general chemical formula (Mg, Fe)$_2$SiO$_4$ where the Mg and Fe endmembers are called forsterite (Mg$_2$SiO$_4$) and fayalite (Fe$_2$SiO$_4$), respectively. Orthopyroxene is (Mg, Fe)SiO$_3$ with Mg and Fe endmembers enstatite (MgSiO$_3$) and ferrosilite (FeSiO$_3$), respectively. At room pressure and temperature, or ambient conditions, forsterite and enstatite have densities of $\sim3200$ kg/m$^3$ \citep[e.g.,][]{Stixrude_2011} whereas fayalite and ferrosilite have ambient densities of $\sim4400$ and 4000 kg/m$^3$, respectively; this means that increasing the amount of Fe relative to Mg in olivine or orthopyroxene will act to increase its density. Between $\sim$14-25 GPa, the upper mantle consists mostly of the olivine polymorphs wadslyite and ringwoodite. The Mg-endmembers of wadslyite and ringwoodite have densities that are $\sim8$ and $\sim10\%$, respectively, larger than forsterite at ambient conditions \citep[e.g.,][]{Stixrude_2011}. The lower mantle begins at pressures $\gtrsim$ 25 GPa, where the orthopyroxene-polymorph silicate-perovskite becomes the dominant mineral, whose Mg-endmember, bridgmanite, is $\sim15\%$ more dense than the Mg-endmember of ringwoodite. Silicate-perovskite remains the dominant stable mineral until $\sim$120 GPa, where it transitions to its high-pressure polymorph, post-perovskite. The zero-pressure densities of the Mg-endmembers of silicate-perovskite and post-perovskite are the same to within $\sim1\%$ and their change in density with increasing pressure are similar as well.

The Earth's core is predominantly Fe, mostly in the liquid phase, and enriched with $\sim$8 wt\% light element(s) \citep[e.g.,][]{McDonough_2003, hirose2021light}. Despite this, most mass-radius-composition models for rocky planets assume that the core is pure Fe in a solid state, often for simplicity but sometimes it is not discussed. Liquid and solid Fe compress at different rates which changes their material density at depth which, all else being equal, will act to change a planet's radius at a given mass. Regardless, assuming the core is entirely solid, there are four known allotropes of solid Fe ($\alpha$, $\beta$, $\gamma$, and $\epsilon$) that are stable over different pressure and temperature ranges like the mantle minerals discussed earlier. However, only the $\epsilon-\mathrm{Fe}$ phase (also known as hexagonal-close-packed Fe or hcp-Fe) is stable above $\sim 100$ GPa, which is exceeded by both the 1$M_\oplus$ and 10$M_\oplus$ planets in Fig. \ref{fig:mantle_section}. 

\section{Common equations of state}
\label{sect:common_eos}
Every material within a planet responds differently to increases in pressure and temperature, meaning each requires a unique, material-dependent EOS to describe its properties at different depths. Since we are concerned with calculating the density, $\rho$, of a material at a given pressure, $P$, and temperature, $T$, within a planet, we limit our scope to EOS of the form $f(\rho, P, T) = 0$. The third-order Birch-Murnaghan \citep[third-order BM or BM3,][]{Birch_1947} and Vinet \citep{Vinet_1987} EOS are the two most commonly used for compressed solids and liquids (discussed in Sect. \ref{sect:BM3} and \ref{sect:vinet_EOS}, respectively). This is likely because both account for the fact that materials become more resistant to compression with increasing pressure, perform well at high pressures, and have historical precedence. Both start from the fact that the pressure of a material under uniform compression is equal to the negative partial derivative of its Helmholtz free energy, $F$, with respect to the volume, $V$, at constant temperature:

\begin{equation}
    P = -\left(\frac{\partial F}{\partial V}\right)_T.
    \label{equ:pressure_helmholtz}
\end{equation}

\subsection{Derivation of the BM3 EOS}
\label{sect:BM3}

The BM3 EOS belongs to the class of BM EOS which use Eulerian finite strain to describe the uniform compression of materials \citep[see Sect. 4.3.1 of ][]{POIRIER_book}. The finite strain ($f$) is the fractional change in a material's density as it is compressed relative to an initial state at some reference $P-T$, expressed as

\begin{equation}
    f = \frac{1}{2}\left[\left(\frac{\rho}{\rho_0}\right)^{2/3} -1\right].
    \label{equ:finit_strain}
\end{equation}

\noindent 
Here, $\rho_0$ is the material's density at the reference $P-T$, usually taken to be ambient conditions. Given that ambient pressure is much smaller than the pressures within rocky planets (e.g., Fig. \ref{fig:mantle_section}), $\rho_0$ is often called the `zero-pressure' density. The Helmholtz free energy (Eq. \ref{equ:pressure_helmholtz}) is approximated as a series expansion of $f$ (Eq. \ref{equ:finit_strain}), where $F = af^2 + bf^3 + cf^4 + ...$ (Eq. 4.27 from \citealp{POIRIER_book}; see also \citealp{Katsura_2019}). Truncating the expansion of $F$ with respect to $f$ to the third order, substituting into Eq. \ref{equ:pressure_helmholtz}, and simplifying gives the BM3 EOS expressed as

\begin{equation}
     P(\rho) = \frac{3}{2}K_0\left[\left(\frac{\rho}{\rho_0} \right)^{\frac{7}{3}} -  \left(\frac{\rho}{\rho_0} \right)^{\frac{5}{3}}\right] \times \left( 1 + \frac{3}{4}(K_0^{'} - 4) \left[\left(\frac{\rho}{\rho_0}\right)^{\frac{2}{3}} - 1 \right]\right),
     \label{equ:BM3}
\end{equation}
where we note that the terms before the $\times$ symbol represent BM2. The $K_0$ term is the bulk modulus at reference conditions, defined as 

\begin{equation}
    \label{equ:k0}
    K_0 \equiv\rho_0\frac{d P}{d \rho}\bigg|_{P=0} = -V_0\frac{d P}{d V}\bigg|_{P=0}\,\,,
\end{equation}

\noindent
and is a measure of the ``incompressibility'' of the material, or how resistant it is to uniform compression. The $K^{'}_0$ term is the first pressure derivative of the bulk modulus at zero-pressure, i.e.,

\begin{equation}
\label{equ:k0prime}
    K_0^{'}\equiv \frac{dK}{dP}\bigg|_{P=0}\,,
\end{equation}
\noindent which dictates how rapidly the effective, or pressure-dependent, bulk modulus $K(P)$ increases with increasing pressure. That is to say, for a given material $\rho_0$ and $K_0$, a low value of $K_0^{'}$ will yield a lower $K(P)$ at a given $P$, and therefore a higher material density, than would be seen for a higher $K_0^{'}$.

\subsection{Derivation of the Vinet EOS}
\label{sect:vinet_EOS}

In contrast to the BM EOS, the Vinet EOS approximates $F$ using an empirical interatomic potential:

\begin{equation}
    F(a) = F_0(1+a)\;e^{-a}\,,
    \label{equ:Vinet_interatomic_potential}
\end{equation}

\noindent where $F_0$ is the equilibrium binding energy of the material and $a = (r - r_0)\,\,/\,\,l$, with $r_0$ and $r$ being the interatomic spacing between the atoms of the material at zero-pressure and high pressures, respectively, and $l$ is a scaling length. By combining Eq. \ref{equ:Vinet_interatomic_potential} with the definitions of $P$, $K_0$, and $K_0^{'}$ (Eq. \ref{equ:pressure_helmholtz}, \ref{equ:k0} and \ref{equ:k0prime}, respectively), it can be shown that $K_0 = F_0\; /\;12\pi l^2r_0$ and $K_0^{'} = 1\; + \;2r_0\:/\:3l$ for the Vinet EOS. Plugging Eq. \ref{equ:Vinet_interatomic_potential} into Eq. \ref{equ:pressure_helmholtz} and simplifying with the substitutions for $K_0$ and $K_0^{'}$, the Vinet EOS is expressed as:

\begin{equation}
     P(\rho) = 3K_0\left(\frac{\rho}{\rho_0}\right)^{\frac{2}{3}}\left[1-\left(\frac{\rho}{\rho_0}\right)^{-\frac{1}{3}}\right]\times \exp \left(\frac{3}{2}(K_0^{'} - 1) \left[1-\left(\frac{\rho}{\rho_0}\right)^{-\frac{1}{3}}\right] \right).
     \label{equ:Vinet}
\end{equation}

\subsection{Comparison of the BM3 and Vinet EOS}
\label{sect:comparison_vinet_BM3}
In practice, the generalized BM3 (Eq. \ref{equ:BM3}) or Vinet (Eq. \ref{equ:Vinet}) EOS are fit to $\rho-P$ data of a material at constant temperature or entropy to
derive the material-dependent isothermal or adiabatic/isentropic EOS parameters ($\rho_0$, $K_0$, and $K_0^{'}$), respectively. For example, if the EOS parameters are fit from isothermal versus isentropic $\rho-P$ data, then $K_0$ is the isothermal bulk modulus, $K_{T0}$, versus the adiabatic bulk modulus, $K_{S0}$, respectively. We note that $K_{T0}\neq K_{S0}$ for a given material, but are relatable to one another through thermodynamic parameters. Further, $K_0$ and its pressure derivatives are specific to the EOS formulation used to fit the data. Namely, Vinet and BM3 EOS fit to the same $\rho-P$ data will yield different $K_0$ and $K_0^{'}$. Therefore, modeling of density at an arbitrary pressure is only valid for the same EOS formulation that was used to derive the EOS from data.

\begin{figure}
    \centering
    \includegraphics[width=0.45\linewidth]{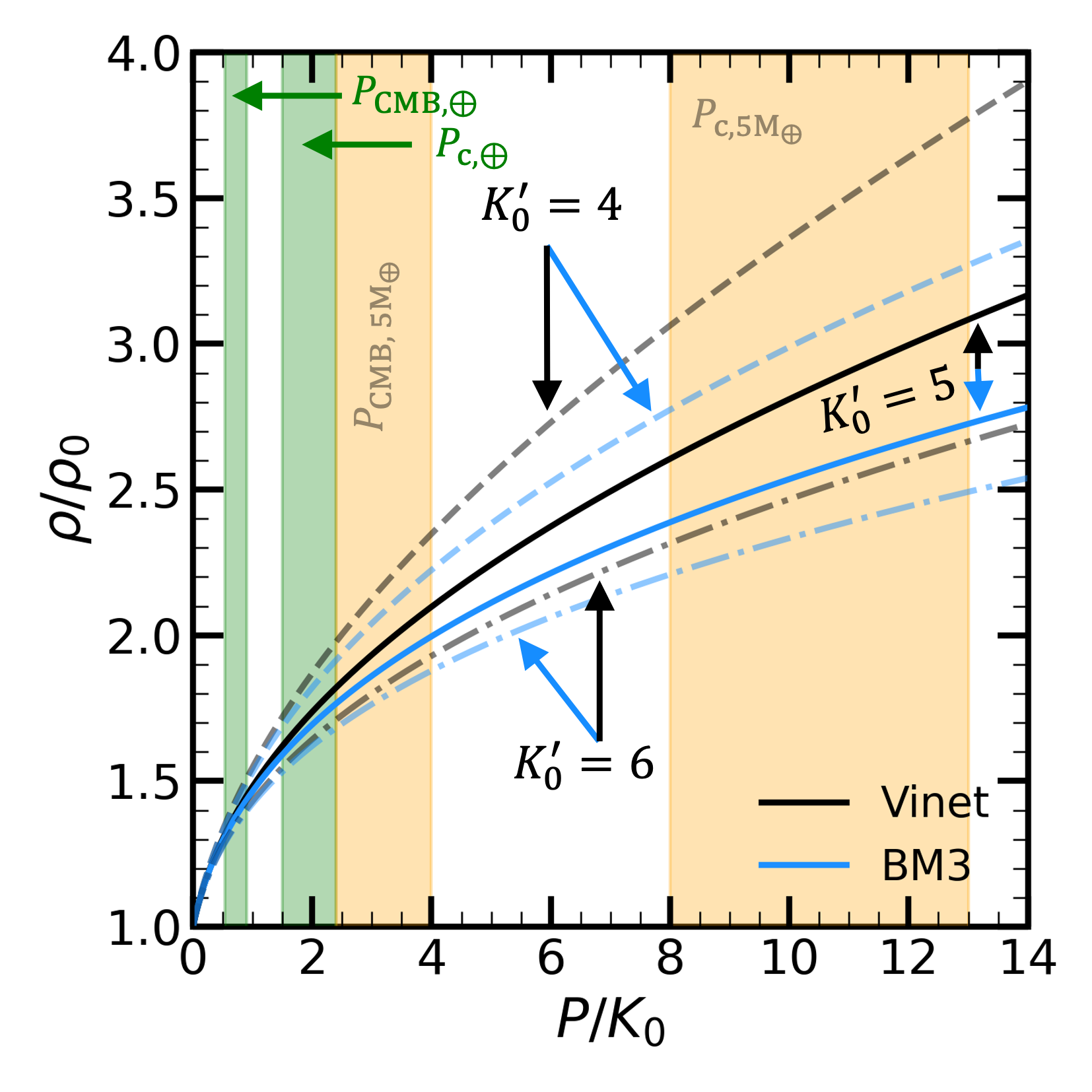}
    \caption{Predicted material densities as a function of pressure for BM3 (blue lines) and Vinet EOS (black lines) for different values of $K_0^{'}$. All densities and pressures are normalized by the zero-pressure density ($\rho_0$) and bulk modulus ($K_0$), respectively. Dashed, solid, and dashed-dotted lines show $K_0^{'} = 4, 5,$ and 6, respectively. The green filled regions show the $P/K_0$ ranges for the core-mantle-boundary (CMB) pressure ($P_\mathrm{CMB, \oplus}$) and central pressure ($P_\mathrm{c, \oplus}$) of Earth for $K_0$ values typical of silicates and Fe. The orange filled regions similarly show $P/K_0$ ranges for the CMB and central pressure of a 5$M_\oplus$ planet ($P_{\mathrm{CMB}, 5M_\oplus}$ and $P_{\mathrm{c}, 5M_\oplus}$, respectively) with Earth-like Fe/Mg.}
    \label{fig:bm3_Vinet_comparison_scaled}
\end{figure}

In Fig. \ref{fig:bm3_Vinet_comparison_scaled}, we compare BM3 (blue lines) and Vinet (black lines) EOS with the same parameters. We show the material density relative to $\rho_0$ as a function of $P$ scaled by $K_0$. In this parameter space, the curves are \textit{independent} of the choice of $\rho_0$ and $K_0$ but not $K_0^{'}$. We therefore show curves with $K_0^{'} = 4,5,$ and 6 (dashed, solid, and dashed-dotted, respectively) which broadly encompasses those for silicates and Fe (e.g., Sect. \ref{sect:selected_EOS_mineral_suites}). Typical values of $\rho_0 \approx 3000-4000\;\mathrm{kg/m^3}$ for silicates and $\sim7000-8500 \mathrm{kg/m^3}$ for liquid and solid Fe. The $K_0$ values for rocky materials are usually between $\sim150-250$ GPa. For a sense of scale, we overlay $P/K_0$ ranges for $K_0=150-250$ GPa at the approximate core-mantle-boundary and central pressures ($P_\mathrm{CMB}$ and $P_\mathrm{c}$, respectively) of Earth (green regions) and a 5$M_\oplus$ Earth-like planet (orange regions). The Earth's $P_\mathrm{CMB,\oplus}=136$ and $P_\mathrm{c, \oplus} = 365$ GPa; whereas, a 5$M_\oplus$ Earth-like planet has a $P_\mathrm{CMB} \sim 600 \;$ and $P_\mathrm{c}\sim 2000$ GPa.

Given that $K_0^{'}$ drives how rapidly the material compressibility changes with pressure, $K_0^{'} = 4$ yields the highest $\rho/\rho_0$ values for all $P/K_0$, while $K_0 = 6$ yields the lowest. The Vinet EOS is more sensitive to changes in $K_0^{'}$ than the BM3 EOS. For a fixed set of EOS parameters within the typical ranges for rocky materials, the BM3 EOS always predicts a lower material density than the Vinet EOS at a given pressure. However, below a $P/K_0 \sim 1$ predict material densities that vary by $\lesssim1\%$. Above this, the two begin to visually deviate. At $P/K_0 = 5$, the BM3 predicts a material density that is $\sim 6.5\%$ lower than that of the Vinet for $K_0^{'} = 4$ and 3.2\% lower for $K_0^{'} = 6$. At $P/K_0 = 10$, the differences increase to $\sim11\%$ and 5.4\% for $K_0^{'} = 4$ and 6, respectively. In the context of rocky exoplanets, EOS with similar $\rho_0$, $K_0$, and $K_0^{'}$ should predict similar material densities within the mantle, as long as $P_\mathrm{CMB}\lesssim P_\mathrm{CMB, \oplus}$. This will not be the case for most core EOS, where differences between the BM3 and Vinet EOS are apparent even at $P_\mathrm{c, \oplus}$. 

We note the above is an overly simple analysis as the EOS parameters were chosen a priori when, in reality, they are fit parameters and will vary between different EOS fit to the same $\rho-P$ data. Both the BM3 and Vinet EOS do equally well at reproducing material density within the pressure ranges used to derive the parameters. Regardless, modeling rocky exoplanets often requires calculating material densities at pressures that are not yet experimentally obtainable. For example, there are many different experimentally-derived EOS for $\epsilon-$Fe used in mass-radius-composition models considered in this work whose \textit{upper} experimental $P$ limits range from $\sim20-1500$ GPa (Sect. \ref{sect:selected_EOS_mineral_suites}), or from well below $P_\mathrm{c, \oplus}$ to the central pressure of a $\sim3-4M_\oplus$ planet. As such, differences in the predicted density of a given material (e.g., $\epsilon-$Fe) between different EOS at the pressures relevant to rocky exoplanets are likely larger than those shown in Fig. \ref{fig:bm3_Vinet_comparison_scaled}. Given that the Vinet EOS does not rely on truncating an expansion at low-order terms, some consider it more robust for extrapolation to high pressures than the BM EOS, but this is still a matter of debate \citep[e.g.,][]{Cohen_2000}.

\subsection{How Thermal Effects are Incorporated into EOS}
\label{sect:thermal_EOS}
While pressure is the primary control on material density, we must also consider how temperature influences the densities of planetary materials. Thermal effects are generally incorporated as a second-order deviation away from the isotherm or isentrope. Perhaps the most common way to account for thermal effects of a material is by adding a thermal pressure correction, $\Delta P_{th}$, to an EOS at some reference pressure, $P_0$, and temperature, $T_0$, such that:

\begin{equation}
    P(\rho,T) = P_0(\rho) + \left[P_{th}(\rho, T) - P_{th}(\rho,T_0)\right] = P_0(\rho) + \Delta P_{th}.
\end{equation}

\noindent
For sense of scale, \citet{Seager_2007} looked at the density reduction from thermal expansion of silicate-perovskite and $\epsilon-$Fe from $T=300-6000$ K. For silicate-perovskite, they find that the reduction in material density is $\lesssim4\%$ at $P>10$ GPa for silicate-perovskite and at $P>100$ GPa for $\epsilon-$Fe. To put it differently, adding a thermal pressure term to an isothermal EOS should only change a material's density by, at most, a few percent at the interior pressure scales of planets (Fig. \ref{fig:mantle_section}).

There are several ways to model $\Delta P_{th}$ and we limit our scope to the commonly used Mie$-$Grüneisen thermal EOS \citep[for derivation see Sect. 2.3.3 of][]{POIRIER_book}, primarily to introduce the standard thermal parameters encountered in mass-radius-composition models. The Mie$-$Grüneisen EOS is written as

\begin{equation}
    \Delta P_{th} = \frac{\gamma}{V}\Delta E_{th},
    \label{equ:mie-gruneisen_eos}
\end{equation}

\noindent where the dimensionless $\gamma$ is called the Grüneisen parameter, $V$ is the material volume, and $\Delta E_{th}$ is the change in thermal energy associated with heating the material from $T_0$ to a higher $T$. The Grüneisen parameter has multiple thermodynamic representations and, therefore, physical interpretations. In practice, it is usually parameterized as a function of $\rho/\rho_0$, the simplest being:

\begin{equation}
    \gamma = \gamma_0\left(\frac{\rho}{\rho_0}\right)^{-q}\,,
    \label{equ:gamma_param}
\end{equation}

\noindent where $\gamma_0$ is the reference Grüneisen parameter and $q$ is a fit parameter. $\Delta E_{th}$ is most commonly described using the Debye model as:

\begin{equation}
    E_{th} = \frac{9nRT}{(\Theta/T)^3}\int^{\Theta/T}_{0}\frac{t^3}{e^t - 1}dt,
    \label{equ:eth_debye}
\end{equation}

\noindent
where $n$ is the number of atoms in the chemical formula of the material (e.g., $n=5$ for MgSiO$_3$ and $n = 7$ for Mg$_2$SiO$_4$), $R$ is the gas constant, and $\Theta$ is the Debye temperature (for a detailed derivation of Eq. \ref{equ:eth_debye}, see Sect. 3 of \citealt{POIRIER_book}). Together, Eq. \ref{equ:mie-gruneisen_eos} and \ref{equ:eth_debye} are often referred to as the Mie$-$Grüneisen-Debye (MGD) EOS. There are various parameterizations for $\Theta$, the simplest being:

\begin{equation}
   \Theta = \Theta_0\left(\frac{\rho}{\rho_0}\right)^{\gamma},
    \label{equ:debye_temp}
\end{equation}

\noindent
where $\Theta_0$ is the reference Debye temperature.

Another approach involves incorporating thermal effects directly into the material density through the thermal expansion coefficient, $\alpha$, often approximated as

\begin{equation}
    \alpha(P,T) = \left(a_0 + a_1T\right)\left(1 + \frac{K_0^{'}}{K_0} P\right)^{-b}\,,
    \label{equ:thermal_expansion_coeff}
\end{equation}

\noindent where $a_0$, $a_1$, and $b$ are fit parameters \citep[e.g.,][]{Fei_1993, Valencia_2007, Magrathea_2022}. In this formulation, the material density is

\begin{equation}
\large
    \rho(P,T) = \rho(P, T_0)\;e^{\left[\int^T_{T_0}\alpha(P,t)dt\right]}.
    \label{equ:alpha_rho_correction}
\end{equation}

\noindent where $\rho(P, T_0)$ is calculated from an isothermal EOS at $T_0$, usually 300 K. Sometimes the thermal expansion coefficient is also incorporated into $K_0$ to give a temperature dependent zero-pressure bulk modulus \citep[e.g.,][]{Sotin_2007}.

\section{Planet Builder}
\label{sect:planet_builder}
Our planet models are based on the ExoPlex mass-radius-composition software \citep{Unterborn23} since it is open-source and has been used in a number of previous works \citep[e.g.,][]{CTU18_Nature_Trappist, Hinkel_Unterborn_2018, unterborn_panero19, schulze_2021, Unterborn23}. The bulk density of a rocky planet is primarily determined by its mass, $M_p$, and the core mass fraction (CMF):

\begin{equation}
    \mathrm{CMF} \equiv \frac{M_{core}}{M_p} = \frac{M_{core}}{M_{core} + M_{mantle}} ,
\label{equ:cmf_definition}
\end{equation}

\noindent where $M_p$ is the total mass of the planet and is equal to the mass of the silicate-dominated mantle, $M_{mantle}$, plus the mass of the Fe-dominated core, $M_{core}$. In the simplest case of a planet with a \textit{pure-Fe} core and \textit{Fe-free} silicate mantle, the mass of the core is 

\begin{equation}
 M_{core} = n_\mathrm{Fe}\times\mu_\mathrm{Fe},
 \label{equ:mass_of_core_in_n}
\end{equation}

\noindent where $n_\mathrm{Fe}$ is the total number of moles of Fe.
Similarly, the mass of the Fe-free mantle can be written in terms of the major and minor mantle elements, assuming they are fully oxidized, as:

\begin{equation}
\begin{split}
    M_{mantle} &= \;(n_\mathrm{Mg}\times\mu_\mathrm{MgO})  \;+ 
    (n_\mathrm{Si}\times\mu_\mathrm{SiO_2}) \;+ \;(n_\mathrm{Ca}\times\mu_\mathrm{CaO}) 
    \;+\;(n_\mathrm{Al}\times\mu_\mathrm{AlO_{1.5}})\\ 
    &= \sum_\mathrm{X = Mg, Si, Ca, Al} n_\mathrm{X}(\mu_X \;+\; n_\mathrm{X,O}\times\mu_\mathrm{O})\,,
    \label{equ:mantle_mass_in_n}  
\end{split}
\end{equation}

\noindent
where $n_\mathrm{X}$ is the number of moles of element $\mathrm{X}$, $\mu_\mathrm{X}$ is the molar mass of element $\mathrm{X}$.
The parameter $n_\mathrm{X,O}$ specifies the number of O atoms per one atom of element $\mathrm{X}$, with $n_\mathrm{X,O} = 1, 2, 1,$ and $1.5$ for or Mg, Si, Ca, and Al, respectively.

Next, we substitute Eqs. \ref{equ:mass_of_core_in_n} and \ref{equ:mantle_mass_in_n} into Eq. \ref{equ:cmf_definition}. We then multiply both the numerator and denominator by $1/n_\mathrm{Mg}$. Since $\mathrm{X/Mg}\equiv n_\mathrm{X}/n_\mathrm{Mg}$, the factors of $1/n_\mathrm{Mg}$ allow the CMF to be expressed in terms of the molar ratios of the major rock-building elements, Fe/Mg and Si/Mg, and the minor mantle elements, Ca/Mg and Al/Mg, such that:

\begin{equation}
    \mathrm{CMF} = \frac{\mathrm{(Fe/Mg)}\mu_\mathrm{Fe}}{\mathrm{(Fe/Mg)}\mu_\mathrm{Fe} + \sum_\mathrm{X = Mg,Si,Ca,Al} \mathrm{(X/Mg)}(\mu_\mathrm{X} + n_\mathrm{X}\mu_\mathrm{O})} = \frac{\mathrm{(Fe/Mg)}\mu_\mathrm{Fe}}{\mathrm{(Fe/Mg)}\mu_\mathrm{Fe} + \bar{\mu}},
    \label{equ:cmf_molar}
\end{equation}

\noindent where

\begin{equation}
    \bar{\mu} \equiv\sum_\mathrm{X = Mg,Si,Ca,Al} \mathrm{(X/Mg)}(\mu_\mathrm{X} + n_\mathrm{X}\mu_\mathrm{O}).
    \label{equ:mu_def}
\end{equation}

\noindent
Given the assumptions made in Eqs. \ref{equ:mass_of_core_in_n}-\ref{equ:mantle_mass_in_n}, Eq. \ref{equ:cmf_molar} is only applicable for rocky planets with pure-Fe cores and Fe-free silicate mantles\footnote{Eqs. \ref{equ:cmf_molar} and \ref{equ:mu_def} are Eqs. 8 and 9 from \citet{Unterborn23}, respectively.}. 

As discussed in Sect. \ref{sect:intro}, there are numerous chemical processes/reactions that may cause a planet to deviate from that of a pure Fe core and Fe-free silicate mantle which, in turn, will alter Eq. \ref{equ:cmf_molar}. \citet{Unterborn23} examine CMF formulations for planets with (1) Fe incorporated into the mantle as FeO, while conserving the planet's bulk Fe/Mg (their Eq. 13); (2) coupled production of FeO in the mantle and light elements in the core (their Eq. 18); and (3) light elements in the core without FeO production (their Section 6.2). For a planet with a non-zero weight fraction of FeO in the mantle, $w_\mathrm{FeO}$, and light elements in the core, $w_\mathrm{LE}$, but whose values are independent of one another (Scenarios 1+3; Sect. \ref{sect:So07} and Sect. \ref{sect:Z16}), the CMF is\footnote{Eq. \ref{equ:cmf_LEs_mant_FeO} is derived by combining Eq. 13 from \citet{Unterborn23} with their expression for a planet with core light elements and no mantle FeO, which is $\mathrm{CMF} = \left[1\;+\;(1-w_\mathrm{LE})\bar{\mu}/\mu_\mathrm{Fe}/(\mathrm{Fe/Mg)}\right]^{-1}$.} :

\begin{equation}
    \mathrm{CMF} = \frac{\mathrm{\mu_{Fe}}\left[\bar{\mu}w_\mathrm{FeO} - \mathrm{\frac{Fe}{Mg}}\mathrm{\mu_{FeO}}\left(1-w_\mathrm{FeO}\right)\right]}{\mathrm{\mu_{Fe}}\left[\bar{\mu}w_\mathrm{FeO} - \mathrm{\frac{Fe}{Mg}}\mathrm{\mu_{FeO}}\left(1-w_\mathrm{FeO}\right)\right] \;- \;\bar{\mu}\mu_\mathrm{FeO}\left(1-w_\mathrm{LE}\right)},
    \label{equ:cmf_LEs_mant_FeO}
\end{equation}

\noindent
For $w_\mathrm{FeO} = w_\mathrm{LE} = 0$, Eq. \ref{equ:cmf_LEs_mant_FeO} naturally reduces to Eq. \ref{equ:cmf_molar}.

Once the planet CMF is calculated, ExoPlex divides the core into $n_{s, core}$ spherical shells, and divides the mass of the core, $M_{core} = \mathrm{CMF}\times M_p$, equally between them. That is, each core shell has a mass of $(\mathrm{CMF}\times M_p)/ n_{s, core}$. Similarly, the mantle is divided into $n_{s, mantle}$ with each having a mass of $M_{mantle}/n_{s, mantle}$, or $(1-\mathrm{CMF})\times M_p\;/\;n_{s, mantle}$. A summation over the mass of all spherical shells is equal to $M_p$. ExoPlex then makes an initial guess for the central pressure of the planet and iteratively solves the following coupled differential radial structure equations in each shell:

\begin{enumerate}
    \item \textbf{The equation of hydrostatic equilibrium} is the balance between self-gravitation and internal pressure within a planet. For a parcel of material at radius $r$ with an infinitesimally small radial thickness, $dr$, the equation of hydrostatic equilibrium is expressed as

\begin{equation}
    \frac{dP(r)}{dr} = -\frac{Gm(r)\:g(r)}{r^2},
    \label{equ:hydro_equilib}
\end{equation}

\noindent where $G$ is the gravitational constant, $m(r)$ is the total mass contained within the sphere of radius $r$, and $\rho(r)$ is the density of the material at $r$.

\item \textbf{The mass within a spherical shell} is classically defined as:

\begin{equation}
    \frac{dm(r)}{dr} = 4 \pi r^2 \rho(r)\,,
\end{equation}

\noindent 
where $dm(r)$ is the infinitesimal mass contained within a shell of thickness $dr$ assuming the planet is radially symmetric.

\noindent
\item \textbf{The thermal EOS} of the constituent materials within the shell is denoted as:

\begin{equation}
    \rho(r) = f(P(r), T(r))\,
    \label{equ:thermal_EOS}
\end{equation}

per Sections \ref{sect:struct_and_minerology_overview} and \ref{sect:common_eos}. Many times there are multiple materials within a given shell, especially within the mantle. When this is the case, the EOS of each material present within the shell is solved for $\rho(P,T)$. The average density of the mixture, $\rho_\mathrm{mix}$, can be calculated using the additive volume rule, i.e., that the volume of the mixture, $V_\mathrm{mix}$, is the sum of the volumes of the individual components at the given $P-T$ conditions \citep[e.g., see][]{Plotnykov20, BICEPS}. For a mixture with $N$ components, 

\begin{equation}
    \rho_{\mathrm{mix}} = 
    \frac{M_\mathrm{mix}}{V_\mathrm{mix}} = 
    \frac{\sum^{N}_{i = 0} m_i}{\sum^{N}_{i = 0}\frac{m_i}{\rho_i}} = 
    \frac{\sum^{N}_{i = 0} w_i\times M_\mathrm{mix}}{\sum^{N}_{i = 0}\frac{w_i\times M_\mathrm{mix}}{\rho_i}} = 
    \left(\sum^{N}_{i = 0}\frac{w_i}{\rho_i}\right)^{-1},
    \label{equ:rho_mix}
\end{equation}

\noindent where $m_i$ is the mass of component $i$ which is equal to its weight fraction, $w_i$, times the total mass of the mixture, $M_\mathrm{mix}$. The sum over all weight fractions must be, by definition, equal to 1, or $\sum^{N}_{i = 0}w_i \equiv 1$. It is sometimes more convenient to express $w_i$ in terms of the molar fraction of component $i$, $x_i$, and its molar mass, $\mu_i$,

\begin{equation}
    w_i = \frac{x_i\mu_i}{\sum^N_{i=0}x_i\mu_i}.
    \label{equ:weight_frac}
\end{equation}

\noindent With this expression, Eq. \ref{equ:rho_mix} becomes:

\begin{equation}
    \rho_\mathrm{mix} = \frac{\sum^{N}_{i = 0}x_i\mu_i}{\sum^{N}_{i = 0}\frac{x_i\mu_i}{\rho_i}}.
    \label{equ:rho_mix_molar}
\end{equation}

\noindent
\item \textbf{The adiabatic temperature profile}, 
\begin{equation}
    \frac{dT(r)}{dr} = -\frac{\gamma(r) \:g(r)\: T(r)}{\Phi(r)} \,,
    \label{equ:temp_gradient}
\end{equation}
\noindent where $\Phi = dP / d\rho$ is the seismic parameter and $T(r)$ is the temperature within the shell. For mixtures, $\gamma$ is assumed to be that of the dominant mineral phase unless otherwise specified in Sect. \ref{sect:selected_EOS_mineral_suites}. Most rocky planet mass-radius-composition models that incorporate temperature assume convection is the primary heat transport mechanism such that the temperature profile of the planet is adiabatic \citep[e.g.,][]{Sotin_2007, Valencia_2007, Dorn15, Dorn_2017}. While not strictly true for Earth, it provides a good approximation \citep[e.g.,][]{turcotte2002geodynamics} and the density of a given material is a much stronger function of pressure than temperature (e.g., Sect. \ref{sect:common_eos}). For models that assume the planet is isothermal \citep[e.g.,][]{Seager_2007}, Eq. \ref{equ:temp_gradient} $\to dT(r) / dr = 0$.

\end{enumerate}

We iteratively solve the interior structure equations until the density within each shell varies by less than 0.001\% from one iteration to the next. Unless otherwise specified, we adopt the ExoPlex boundary conditions of $P(R_p) = 0.0001 \;\mathrm{GPa} = 1 \;\mathrm{bar}\simeq 1\;\mathrm{atm}$ and $T(R_p) = T_\mathrm{pot} = 1600$ K, where $T_\mathrm{pot}$ is called the mantle potential temperature and is defined as the temperature a parcel of solid mantle material would have at the surface assuming no melting occurs. By default, ExoPlex sets $n_{s, core} = 600$ and $n_{s, mantle} = 500$, which we adopt as well, since they were selected primarily to (1) accurately capture the complicated upper mantle phase transitions, (2) accurately model the pressure gradient within the core, which will be greater than that of the mantle (e.g., Eq. \ref{equ:hydro_equilib}), and (3) ensure convergence, i.e., the calculated radius and density do not change measurably when increasing the number of shells. If point (3) is satisfied, (1) and (2) should be as well. For a 1$M_\oplus$ planet with an Earth-like composition and fixed $n_\mathrm{s, core}=600$, the calculated bulk planet density varies by $<0.005\%$ per each additional 50 shells above $n_\mathrm{s, mant} = 500$. Similarly, for fixed $n_\mathrm{s, mant} = 500$, the difference in bulk planet density is $<0.002\%$ per each additional 50 shells above $n_\mathrm{s, core}=600$.

\makeatletter
\newcommand\footnoteref[1]{\protected@xdef\@thefnmark{\ref{#1}}\@footnotemark}
\makeatother

\section{Selected EOS+mineral suites for rocky planets}
\label{sect:selected_EOS_mineral_suites}

 We investigate the EOS+mineral suites underlying the following mass-radius-composition models (Table \ref{table:EOS_suites}): \citet[][or S07]{Seager_2007}, \citet[][or So07]{Sotin_2007}, \citet[][or V07]{Valencia_2007}, \citet[][or ZS13]{Zeng_Sasselov_2013}, \citet[][or D15]{Dorn15}, \citet[][or Z16]{Zeng16_emperical_MR_function}, \citet[][or D17]{dorn17_bayes}, MAGRATHEA \citep{Magrathea_2022}, and ExoPlex \citep{Unterborn23}. The models were selected because they encompass a wide range of material EOS (Sect. \ref{sect:common_eos}), assumptions about the materials/minerals present within the planet (Sect. \ref{fig:mantle_section}), and are widely used in the literature focused on small planets. While not exhaustive\footnote{We recognize the excellent work done by \citet{Zeng2019, Plotnykov20, Adibekyan_2021, Adibekyan_2024, BICEPS}, but we opted not to include them in our analysis for a variety of reasons.}, our list provides a good proxy for the differences between the suites used in small planet mass-radius-composition models.

We reiterate that we are investigating the EOS+mineral suites underlying the mass-radius-composition models, not the mass-radius-composition models themselves. This is a subtle but important distinction since, in all but two cases (ExoPlex and MAGRATHEA), the mass-radius-composition models are not open source and cannot be run directly. Instead, we apply the suites to the planet builder outlined in Sect. \ref{sect:planet_builder}. Doing so ensures that we are comparing the suites self-consistently, rather than using different methods for solving the interior structure equations (Eq. \ref{equ:hydro_equilib}-\ref{equ:temp_gradient}). Our study requires interpreting and applying the methods from papers across $\sim2$ decades (Table \ref{table:EOS_suites}), who included various levels of detail regarding their analysis; any potential misinterpretations or misunderstandings in our analysis are unintentional.

\begin{table}
    \centering

    \scriptsize
    \begin{tabular}{|c|c|c|c|c|c|c|c|}
        
    \hline
         \textbf{EOS+mineral suite} & \textbf{Core/Mantle} & \textbf{Material} & \textbf{EOS} & $\pmb{\rho_0}$ & $\pmb{K_0}$ & $\pmb{K_0^{'}}$ & \textbf{EOS} \\
         & & & \textbf{Type} & [kg/m$^3$]& [GPa] & & \textbf{Refs}\\\hline
         \hline
         
         \citet{Seager_2007} & Core & $\epsilon$-Fe & Vinet & 8300 & 156.2 & 6.08 & \citet{Anderson_2001}\\\cline{2-8}
         (S07) & Mantle & bridgmanite & BM4\footnote{$K_0^{''} = -0.016$ GPa$^{-1}$} & 4100 & 247 & 3.97 &  \citet{Karki_2000}\\\hline
         \hline
         
         \citet{Sotin_2007} & Core & $\epsilon-\mathrm{Fe}_{0.87}\,\mathrm{FeS}_{0.13}$ & BM3 & 7357 & 135 & 6 & \citet{Uchida_2001} \\
        (So07) & & & & & & & \citet{Kavner_2001} \\\cline{2-8}
        & Upper Mantle & olivine & BM3 & 3347 & 128 & 4.3  & \citet{duffy1995elasticity} \\
        & & & & & & &  \citet{Bouhifd_1996} \\\cline{3-8}
        & & orthopyroxene & BM3 & 3299 &  111 & 7 & \citet{VACHER1998275} \\\cline{2-8}
         & Lower Mantle & silicate-perovskite & BM3 & 4219 & 263 & 3.9 & \citet{hemley1992constraints}\\\cline{3-7}
         & & magnesiow{\"u}stite & BM3 & 3830 & 157 & 4.4 & \\\hline
         \hline

         \citet{Valencia_2007} & Core & $\epsilon-$Fe & Vinet & 8300 & 160.2 & 5.82 & \citet{Williams_1997}\\\cline{3-7}
         (V07) & & $\epsilon-\mathrm{Fe}_{0.8}\,\mathrm{FeS}_{0.2}$  & Vinet & 7171 & 150.2 & 5.675 & \citet{Uchida_2001}\\\cline{2-8}
         & Upper Mantle & olivine & Vinet & 3347 & 126.8 & 4.274 & \citet{Stixrude_2005} \\\cline{3-7} 
         & & ringwoodite & Vinet & 3644 & 174.5 & 4.274 & \\
         & & + wadsleyite & & & & & \\\cline{2-7}
         & Lower Mantle & silicate-perovskite & Vinet & 4152 & 223.6 & 4.274 &\\
         & & +magnesiowüstite & & & & & \\\cline{3-8}
         & & post-perovskite & Vinet & 4270 & 233.6 & 4.524 & \citet{Tsuchiya_2004}\\
         & & +magnesiowüstite & & & & &\citet{Stixrude_2005} \\\hline
         \hline
         \citet{Zeng_Sasselov_2013} & Core & $\epsilon-\mathrm{Fe}$ & Vinet & 8300 & 156.2 & 6.08 & \citet{Anderson_2001} \\\cline{2-8}
         (ZS13) & Mantle &  bridgmanite & BM3 & 4087 & 232 & 3.86 & \citet{Caracas_2005a}\\\cline{3-8}
         & & post-perovskite & BM3 & 4080 & 203.0 & 4.19 & \citet{Caracas_Cohen_2008} \\\hline
         \hline
         \citet{Zeng16_emperical_MR_function} & Core & Earth-like & BM2 & 7050 & 201.0 & & \citet{PREM} \\\cline{2-8}
         (Z16) & Upper Mantle & Earth-like & Lin. Interp. & & & & \citet{Stacy_Davis_2009}\\\cline{2-8}
         & Lower Mantle & Earth-like &  BM2 & 3980 & 206 & &\citet{PREM} \\\hline
         \hline
         \citet{Dorn15} & Core & $\epsilon-$Fe & BM3 & 8341 & 173.98 & 5.297 & \citet{belonoshko_2010} \\\cline{2-8}
         (D15) & Mantle & Many\footnote{\label{perplex_note} This EOS+mineral suite uses the \texttt{Perple\_X} thermodynamic
         equilibrium software \citep{connolly_2009} to determine the mineral assemblage within each mantle shell that is thermodynamically consistent with the bulk mantle composition thereby developing a self-consistent upper and lower mantle. } & BM3 & & & & \citet{Stixrude_2011}\\\hline
         \hline
         \citet{dorn17_bayes} & Core & $\epsilon-$Fe & BM3 & 8878 & 253.844 & 4.719 & \citet{Bouchet_2013} \\\cline{2-8}
         (D17) & Mantle & Many\textsuperscript{\ref{perplex_note}} & BM3 & & & &  \citet{Stixrude_2011} \\\hline
         \hline
         \citet{Huang2018} & Core & $\epsilon-$Fe & Vinet & 8430 & 177.7 & 5.64 & \citet{Smith18_Fe_EoS} \\\cline{2-8}
         MAGRATHEA v1.0 & Mantle & bridgmanite & Vinet & 3983 & 230.05 & 4.142 & \citet{Oganov_2004} \\
         & & & & & & & \citet{ONO2005914} \\\cline{3-8}
         & & post-perovskite & Keane & 4059.5 & 203.0 & 5.35 & \citet{Sakai_2016} \\\hline
         \hline
         \citet{Unterborn23} &  Core & liquid-Fe & BM4\footnote{$K_0^{''} = -0.043$ GPa$^{-1}$} & 7019.0 & 109.7 & 4.661 & \citet{anderson1994equation} \\\cline{2-8}
         ExoPlex & Mantle & Many\textsuperscript{\ref{perplex_note}}& BM3 & & & &\citet{Stixrude_2011}\\\hline

    \end{tabular}

    \caption{Summary of the EOS+mineral suites in our analysis, specifically their respective EOS and isothermal/isentropic parameters as well as the included materials. }
    \label{table:EOS_suites}
\end{table}

\subsection{Seager+ 2007 (S07)}
\label{sect:S07}

\citet[][or S07]{Seager_2007} investigates a few different EOS for various core and mantle materials, but we limit our scope to the EOS+mineral suite that we interpret as being the `default' or `preferred' suite, specifically as outlined in detail in their Sect. 3.3. The default S07 suite assumes a pure-Fe core and a Fe-free silicate mantle with no minor mantle elements. The planet CMF as a function of Fe/Mg can therefore be calculated via Eq. \ref{equ:cmf_molar}, where Si/Mg = 1, Ca/Mg = 0, and Al/Mg = 0. The authors implement isothermal EOS for both core and mantle materials. The core is assumed to be pure $\epsilon$-Fe as described by a Vinet EOS (Sect. \ref{sect:vinet_EOS}), specifically as presented in \citet{Anderson_2001} which was fit to experimental data at $T=300$ K and $P\leq 330$ GPa. S07 uses this EOS for $P\leq20.9$ TPa, where they switch to a Thomas-Fermi-Dirac (TFD) theoretical EOS \citep{Salpeter_Zapolsky}. When replicating their EOS+mineral suite within the planet builder (Sect. \ref{sect:planet_builder}), we do not implement the TFD EOS as the central pressures of small planets are limited to several TPa \citep{unterborn_panero19}. S07 does not include any upper mantle minerals, assuming the mantle is entirely bridgmanite (or MgSiO$_3$ silicate-perovskite) for $P\leq 13.5$ TPa. The authors adopt the 300 K isothermal BM4 EOS for bridgmanite from \citet{Karki_2000} derived from density functional calculations for $P\leq150$ GPa. For $P>13.5$ TPa, the authors again switch to a theoretical TFD EOS which we do not implement as the $P_\mathrm{CMB}$ of small planets are limited to $\lesssim$1-2 TPa \citep[e.g.,][]{unterborn_panero19}.

\subsection{Sotin+ 2007 (So07)}
\label{sect:So07}

\citet[][or So07]{Sotin_2007} assumes the planet core is a mixture of 87 mol \% Fe and 13 mol \% FeS, or Fe$_{0.87}$FeS$_{0.13}$. For the pure-Fe component of the mixture, the authors adopt an isothermal BM3 EOS (Sect. \ref{sect:BM3}) and a MGD thermal EOS (Sect. \ref{sect:thermal_EOS}) for $\epsilon-$Fe from \citet{Uchida_2001}, whose experiments covered pressures $\leq$20 GPa and temperatures $\leq$1500 K. So07 cites \citet{Kavner_2001} for the $\rho_0$ of FeS. While not explicitly stated, we assume that So07 adopts an average $\rho_0$ for the Fe$_{0.87}$FeS$_{0.13}$ mixture but otherwise uses the $\epsilon-$Fe EOS parameters. Using Eq. \ref{equ:rho_mix_molar} with the quoted $\rho_0 = 8340$ and $4900$ kg/m$^3$ for $\epsilon-$Fe and FeS, respectively, we calculate a $\rho_0 = 7365.5$ kg/m$^3$ for Fe$_{0.87}$FeS$_{0.13}$.

So07 divides the mantle into an upper mantle composed of a mixture of olivine + orthopyroxene\footnote{So07 calls this as enstatite but it is a mixture of enstatite and ferrosilite. We note it is not uncommon for such a mixture to be labeled as enstatite if the ferrosilite content is $< 50$ mole \%. However, we use the term enstatite specifically for the Mg-endmember of the orthpyroxene enstatite-ferrosilite series.} and a lower mantle of silicate-perovskite + magnesiow{\"u}stite. The general chemical formula for magnesiow{\"u}stite is (Mg, Fe)O, where the Mg- and Fe-endmembers are periclase and w{\"u}stite, respectively. The authors assume that olivine + orthopyroxene transitions to silicate-perovskite + magnesiow{\"u}stite when $P \geq \mathrm{25 GPa} - 0.0017\times(T- \mathrm{800 K})$, per \citet{IRIFUNE1987324}. Adopting the conventions of So07, the upper mantle is made of $x_3$ by mole of olivine and $1-x_3$ by mole of orthopyroxene. These minerals are then comprised of $y_3$ and $1-y_3$ mole fraction of their Fe and Mg endmembers, respectively. Put differently, a mole of material in the upper mantle contains a fraction of $x_3\times(1-y_3)$ of forsterite. Similarly, the lower mantle contains $x_2$ and $1-x_2$ mole fractions of silicate-perovskite and magnesiow{\"u}stite, respectively, comprised of $1-y_2$ of their Mg-endmembers and $y_2$ of their Fe-endmembers. The general chemical formula for the upper mantle is then

\begin{equation}
      x_3\underbrace{\big[(\mathrm{Mg}_{2(1-y_3)}, \mathrm{Fe}_{2y_3})\mathrm{SiO}_4\big]}_\text{olivine} \; + \; (1-x_3)\underbrace{\big[\mathrm{Mg}_{2(1-y_3)}, \mathrm{Fe}_{2y_3}\mathrm{Si_2 O_6} \big]}_\text{orthopyroxene},
      \label{equ:so07_UM} 
\end{equation}

\noindent and the lower mantle is 

\begin{equation}
           x_2\underbrace{\big[(\mathrm{Mg}_{1-y_2},\mathrm{Fe}_{y_2})\mathrm{SiO}_3 \big]}_\text{silicate-perovskite} \;+ \;(1-x_2)\underbrace{\big[(\mathrm{Mg}_{1-y_2},\mathrm{Fe}_{y_2})\mathrm{O}\big]}_\text{magnesiow{\"u}stite}.
           \label{equ:so07_LM}
\end{equation}

\noindent 
So07 further assumes the mantle is assumed to be chemically homogeneous (see their Sect. 2.2), which adds the constraints that $y_2 = y_3$ and $x_2 = 1 - x_3/2$. For this study, we use the Earth-like ``Model 1'' of So07 (see their Table 2) with Si/Mg = 0.88 and a molar Mg/(Mg+Fe) fraction in the mantle of 0.9. This choice of Si/Mg and Mg/(Mg+Fe) gives $y_2 = y_3 = 0.1$, $x_2 = 0.8$, and $x_3=0.4$. It is straightforward to use Eq. \ref{equ:so07_UM} or \ref{equ:so07_LM} to show that the $w_\mathrm{FeO}$ for the mantle is

\begin{equation}
   w_\mathrm{FeO} = \frac{y_2\mu_\mathrm{FeO}}{y_2\mu_\mathrm{FeO} + (1-y_2)\mu_\mathrm{MgO} + x_2\mu_\mathrm{SiO_2}},
\end{equation}

\noindent where the denominator is a modified version of Eq. \ref{equ:mantle_mass_in_n}. For $x_2 = 0.8$ and $y_2=0.1$, $w_\mathrm{FeO} \simeq 0.08$. Similarly, the weight fraction of light elements in the core is  $w_\mathrm{LE} = 0.13\mu_\mathrm{S}/(0.13\mu_\mathrm{S} + \mu_\mathrm{Fe}) \simeq 0.07$, where $\mu_\mathrm{S} = 32.065$ g/mol is the molar mass of S. The planet CMF for an input Fe/Mg can then be found via Eq. \ref{equ:cmf_LEs_mant_FeO} with $w_\mathrm{FeO} = 0.08$, $w_\mathrm{LE} = 0.07$, Si/Mg = 0.88, Ca/Mg = 0.0, and Al/Mg = 0.0.

So07 use a BM3 EOS for all mantle minerals, but incorporate the thermal effect in two different ways (see their Appendix A). The isothermal and thermal parameters for silicate-perovskite and magnesiow{\"u}stite are those derived by \citet{hemley1992constraints} who fit to experimental data ranging from 0-30 GPa and 300-900K. The chosen olivine EOS parameters originate from the experiments of \citet{duffy1995elasticity}, which spans 3-16 GPa, and \citet{Bouhifd_1996}, whose isobaric analysis had a range of 400-2160 K\footnote{The authors use the values listed in Table 2 and 3 of \citet{VACHER1998275}, who themselves cite \citet{duffy1995elasticity} for the bulk modulus and its pressure derivative as well as \citet{Bouhifd_1996} for the thermal expansion coefficients. The temperature derivative of the bulk modulus listed in Table 1 of So07 is likely taken from \citet{Isaak_1992}.}. For the adopted orthopyroxene EOS parameters, the authors cite \citet{VACHER1998275} which reviews a number of experiments\footnote{So07 also references \citet{anderson1991thermoelastic} with respect to orthopyroxene, but the relationship is not immediately clear since \citet{anderson1991thermoelastic} does not study orthopyroxene.}. The isothermal parameters are from \citet{angel1994equations} whose experimental data covers pressures up to $\sim$8.5 GPa, and the thermal parameters are from \citet{Zhao_1995} whose P-V-T data spans $P\leq$4.6 GPa and $T\leq$ 1000 K. Similar to the core, we calculate $\rho_0$ values for olivine, orthopyroxene, silicate-perovskite, and magnesiow{\"u}stite by averaging the $\rho_0$ values of their respective Fe- and Mg-endmembers according to Eq. \ref{equ:rho_mix}-\ref{equ:rho_mix_molar} and the relevant $y_2$ and $y_3$ values discussed above, but otherwise use the EOS parameters of the Mg-endmember phases. The EOS parameters for these four minerals are listed in Table \ref{table:EOS_suites}. Within each upper mantle layer, we first determine the densities of olivine and orthopyroxene from their respective EOS. We then compute the density of the mixture using Eq. \ref{equ:rho_mix}-\ref{equ:rho_mix_molar} with $x_3=0.4$. We follow the same procedure to determine the density of the silicate-perovskite and magnesiow{\"u}stite mixture within the lower mantle layers with $x_2 = 0.8$. The thermal EOS parameters are assumed to be that of the dominant mineral phase within each layer.

\subsection{Valencia+ 2007 (V07)}
\label{sect:V07}

\citet[][or V07]{Valencia_2007} uses a Vinet isothermal EOS with a MGD thermal EOS for all rocky materials. Where applicable, V07 refit isothermal data from a variety of sources to a Vinet EOS to determine a new set of isothermal EOS parameters for each material. While we summarize the data sources referenced in V07, we did not attempt to reproduce their data fitting. V07 considers two cases for the Fe-dominated core: (1) pure $\epsilon-$Fe, and (2) a mixture of 80 mol \% $\epsilon-$Fe and 20 mol \% FeS, or Fe$_{0.8}$(FeS)$_{0.2}$. For the isothermal EOS parameters of all core materials, V07 cites \citet{WILLIAMS199749}, which tabulates parameters from a few other sources. The isothermal EOS parameters for $\epsilon-$Fe originate from \citet{Mao_1990}, which measured $P-V$ data for $\epsilon-$Fe up to 300 GPa. For FeS, V07 is likely refitting data from \citet{Ahrens_1979}. We are able to reproduce the quoted $\rho_0$ for the Fe$_{0.8}$(FeS)$_{0.2}$ mixture in V07 (their Table 1) using Eq. \ref{equ:rho_mix_molar} with the $\rho_0$ values for $\epsilon-$Fe and FeS. The thermal EOS parameters for both core cases are that of the dominant $\epsilon-$Fe phase and are taken from \citet{Uchida_2001}, the same as in So07. 

V07 splits the mantle into an upper and lower mantle with intermediate phase transitions in both regions. The upper mantle is composed of olivine which transitions to a mixture of ringwoodite+wadsleyite at $T$ (K) $= 400P$ (GPa) $- 4287$. Given olivine, wadsleyite, and ringwoodite are polymorphs of the same material, the chemical formula for the upper mantle is simply:

\begin{equation}
    \underbrace{(\mathrm{Mg}_{1-y_\mathrm{1}}, \mathrm{Fe}_{y_1})\mathrm{SiO_4}}_\mathrm{olivine/wadsleyite/ringwoodite}\,,
    \label{equ:valencia_upper_mant}
\end{equation}

\noindent where $y_1$ is the molar fraction of the Fe-endmember (FeSiO$_4$). The lower mantle begins when wadsleyite+ringwoodite transition to a mixture of silicate-perovskite+magnesiowüstite\footnote{Listed as ferromagnesiowüstite in V07}. The phase transition is given by $P = 22.6$ GPa if $T > 1750$ K. Otherwise, it is $T$ (K) = $13,573 - 500$P (GPa). Finally, the authors assume silicate-perovskite transitions to post-perovskite when $T = 133P - 13,875$\footnote{We believe the offset term for the pv$\to$ppv transition should read 13875 K, not 1392 K, in Table 2 of \citet{Valencia_2007} per \citet{Tsuchiya_2004}}, to create a base mantle layer of post-perovskite+magnesiowüstite. For both regions of the lower mantle, magnesiowüstite has a molar fraction of $x_\mathrm{mw} = 0.3$. V07 assumes that silicate-perovskite and magnesiowüstite consist of the same $y_1$ mole fraction of their respective Fe-endmembers as the upper mantle minerals. This is not true for post-perovskite, which is assumed to have a different molar fraction of Fe within its mineral structure, $y_2$. The general chemical formula for the lower mantle is then:

\begin{equation}     
    \begin{cases}
       (1-x_\mathrm{mw})\underbrace{\left[(\mathrm{Mg}_{1-y_1}, \mathrm{Fe}_{y_1})\mathrm{SiO_3}\right]}_{\mathrm{silicate-perovskite}} + \; x_\mathrm{mw}\underbrace{\left[\mathrm{Mg}_{1-y_1}, \mathrm{Fe}_{y_1})\mathrm{O}\right]}_\text{magnesiowüstite} ,\;\;\;\mathrm{ if }\;\; T < 133P - 13875, \,\textrm{and} \\
       (1-x_\mathrm{mw})\underbrace{\left[(\mathrm{Mg}_{1-y_2}, \mathrm{Fe}_{y_2})\mathrm{SiO_3}\right]}_\mathrm{post-perovskite} + \;x_\mathrm{mw}\underbrace{\left[(\mathrm{Mg}_{1-y_1}, \mathrm{Fe}_{y_1})\mathrm{O}\right]}_\text{magnesiowüstite} ,\;\;\:\text{otherwise}.
    \end{cases}  
\end{equation}

\noindent
All mantle minerals in V07 are assumed to have 10 mol \% Fe except for post-perovskite which is assigned 20 mol \% Fe, or $y_1 = 0.1$ and $y_2=0.2$. This choice acts to create a \textit{mass-dependent} CMF for fixed planet Fe/Mg given that the mantle volume fraction of post-perovskite increases with planet mass (Sect. \ref{sect:struct_and_minerology_overview}). As such, there is no closed-form relationship between CMF and Fe/Mg like Eq. \ref{equ:cmf_molar} or \ref{equ:cmf_LEs_mant_FeO}. Instead, we vary the planet CMF until the input Fe/Mg is reproduced. 

With the exception of post-perovskite, \citet{Stixrude_2005} is cited as the source of the EOS parameters of all mantle minerals which, in turn, pulls from many sources. The EOS parameters for post-perovskite are derived from \citet{Tsuchiya_2004}, who studied the properties from first-principle calculations up to $\sim$200 GPa and 4000 K. V07 states that the relevant thermal EOS parameters $\theta_0$, $\gamma_0$, and $q$ are taken from the dominant phase for each material (e.g., the thermal EOS parameters for olivine are taken from forsterite). We find that this statement is true for all materials/mixtures within their models except for the ringwoodite+wadsleyite layer, where the $\Theta_0$ parameter is that of Mg-ringwoodite, but $q$ is that of Mg-wadsleyite. The $\gamma_0$ parameter for ringwoodite+wadsleyite is also likely that of Mg-wadsleyite as the values listed in Table 1 of V07 and Table A1 of \citet{Stixrude_2005} are similar, but not identical.

\subsection{Zeng \& Sasselov 2013 (ZS13)}
\label{sect:ZS13}
\citet[][or ZS13]{Zeng_Sasselov_2013} adopts isothermal EOS for all rocky materials in their models. The core is assumed to be pure $\epsilon-$Fe, described by the same Vinet EOS as S07 \citep[][or Sect. \ref{sect:S07}]{Anderson_2001}. The silicate mantle is Fe-free with Si/Mg = 1 and no minor mantle elements (Ca/Mg = Al/Mg = 0). ZS13 does not include any upper mantle minerals in their models. They assume the lowest-pressure mantle phase is bridgmanite and that it transitions to post-perovskite at 122 GPa. The authors point to \citet{Caracas_Cohen_2008} as the source of the EOS parameters for both bridgmanite and post-perovskite, who studied the properties of post-perovskite up to 180 GPa via density functional theory \citep[or DFT, see e.g.,][]{DFT_book} calculations up to 180 GPa. \citet{Caracas_Cohen_2008} state they calculate the properties of bridgmanite, but do not detail the results; however, they include a figure comparing the post-perovskite phase with an earlier work on silicate-perovskite \citep{Caracas_2005a} up to 180 GPa, which we assume are the source of the parameters for bridgmanite used in ZS13.  

Finally, ZS13 uses a dissociation of MgSiO$_3$ $\to$ MgO+MgSi$_2$O$_5$ above 900 GPa, and then a further dissociation of MgSi$_2$O$_5$ $\to$ MgO+SiO$_2$ at 2.1 TPa according to the calculations of \citet{UMEMOTO2011225}. We do not incorporate either dissociation as \citet{UMEMOTO2011225} provides only the relevant thermodynamic properties for MgSi$_2$O$_5$ (their Table 2) and we are unable to locate the parameters for the remaining materials. Given the core-mantle-boundary pressures of rocky planets are limited to $\sim1$ TPa \citep[][]{unterborn_panero19}, excluding these dissociations should not affect our models appreciably as the MgSiO$_3$ $\to$ MgO+MgSi$_2$O$_5$ will only be relevant for the upper limit of our mass range. Therefore, we do not expect the MgSi$_2$O$_5$ $\to$ MgO+SiO$_2$ to be relevant to the planets considered here.

\subsection{Zeng+ 2016 (Z16)}
\label{sect:Z16}

\citet[][or Z16]{Zeng16_emperical_MR_function} uses seismologically-derived P-$\rho$ data of the Earth \citep[PREM: Preliminary Reference Earth Model,][]{PREM} as the basis for their mass-radius-composition model. The authors present BM2 EOS fits for both the Earth's inner solid core and outer liquid core, but only incorporate that of the outer core into their planet models. As such, ZS13 is assuming the cores of small planets are liquid Fe with an Earth-like light element enrichment. The authors justify this choice by noting that the solid inner core makes up a small volume fraction of Earth and may be non-existent in more massive planets given their higher interior temperatures and slower cooling rates. Above 12 TPa, the authors switch from the BM2 EOS to a theoretical EOS for Fe. We do not incorporate the theoretical EOS as the central pressures of small planets are limited to several TPa. For the upper mantle, Z16 employs a linear interpolation of PREM-based data in \citet{Stacy_Davis_2009} up to 23.83 GPa. Above this pressure, the authors switch to a BM2 fit of PREM lower-mantle data. Finally, above 3.5 TPa, Z16 switch to a theoretical EOS for MgSiO$_3$. We do not incorporate the TFD EOS for either the core or mantle for the same reasons as in S07 (or Sect. \ref{sect:S07}): the core-mantle-boundary pressure of the planets in this work are limited to a few and $\lesssim1-2$ TPa, respectively.

Unlike the other mass-radius-composition models outlined so far, by fitting directly to PREM, the mantle and core materials are not explicitly defined but implicitly assumed to have both Earth-like compositional and thermal structure. This makes it difficult to write a well-defined function relating CMF to the molar ratios of the rock-building elements. As such, we use Eq. \ref{equ:cmf_LEs_mant_FeO} and adopt Earth-like values of $w_\mathrm{LE} = 0.08$, $w_\mathrm{FeO} = 0.08$, Si/Mg = 0.9, Ca/Mg = 0.07, and Al/Mg = 0.09 \citep[e.g.,][]{McDonough_2003, Unterborn23}.

\subsection{MAGRATHEA v1.0}
\label{sect:magrathea}

The open-source MAGRATHEA code\footnote{MAGRATHEA GitHub: \href{https://github.com/Huang-CL/Magrathea}{https://github.com/Huang-CL/Magrathea}} is highly modular, containing many EOS for Fe and silicates. For the purposes of this study, we adopt the default EOS+mineral suite outlined in \citet{Magrathea_2022} and refer to it as MAGRATHEA v1.0 hereafter to distinguish it from future versions of the MAGRATHEA software \citep[e.g.,][]{rice2025magratheav2planetaryinterior}. The core is assumed to be pure $\epsilon-$Fe described by an Vinet EOS and a MGD thermal EOS. The isothermal parameters and $\theta_0$ are those of \citet{Smith18_Fe_EoS}, derived from isentropic P-$\rho$ data up to 1.4 GPa, and the $\gamma$ parameter used by MAGRATHEA v1.0 is derived by refitting the $\gamma-\rho$ data \citep[Fig. 3b of][]{Smith18_Fe_EoS} with a different formulation of the gr{\"u}neisen parameter\footnote{See Eq. 13 of \citet{Magrathea_2022} for the correct form of the gr{\"u}neisen parameter.}.

MAGRATHEA v1.0 assumes that the mantle is pure MgSiO$_3$ and does not include any upper mantle mineral phases\footnote{The current GitHub version \textbf{(v2.0.0-beta)} of MAGRATHEA does appear to include upper mantle minerals as a default, but we base our analysis on what is published in \citet{Magrathea_2022}}. The lowest pressure mantle phase is bridgmanite (i.e., MgSiO$_3$ silicate-perovskite) which transitions to MgSiO$_3$ post-perovskite at P (GPa) $= 130 + 0.007(T-2500$K) \citep{ONO2005914}. The authors use a Vinet + MGD EOS for bridgmanite, where the isothermal parameters come from  theoretical calculations of $P \sim 20-100$ GPa \citep[per Fig. 2 of][]{Oganov_2004}. The thermal parameters are likely from \citet{ONO2005914}, which experimentally observed the bridgmanite phase over $P=112.5-133.3$ GPa and $T=1900-3350$ K (per their Table 5). Finally, MAGRATHEA v1.0 adopts an isothermal Keane EOS \citep{keane1954investigation, STACEY2004137} and thermal MGD EOS for MgSiO$_3$ post-perovskite, with the parameters of \citet{Sakai_2016} derived from experimental P-V-T data over $P\sim118-265$ GPa and $T\sim 300-2710$ K.

\subsection{ExoPlex}
\label{subsection:exoplex}

The open-source ExoPlex software\footnote{The ExoPlex GitHub: \href{https://github.com/CaymanUnterborn/ExoPlex}{https://github.com/CaymanUnterborn/ExoPlex}} allows variable light elements in the core (O, Si, or S) and variable mantle minerology, including FeO and minor mantle elements. With the exception of Fe/Mg, we adopt the default ExoPlex settings for this work: $w_\mathrm{FeO} = 0$, $w_\mathrm{LE} = 0$, Si/Mg = 0.9, Ca/Mg = 0.07, and Al/Mg = 0.09. The relationship between CMF and bulk Fe/Mg is then given via Eq. \ref{equ:cmf_molar}. The core is assumed to be liquid Fe described by the isentropic BM4 EOS of \citet{anderson1994equation}, derived from experimental data up to 1 TPa, and is applied using the BurnMan thermodynamic and geophysics toolkit \citep[][]{cottaar_burnman, myhill2023burnman}. BurnMan adopts the reference isentrope EOS parameters (i.e., $\rho_{S0}$, $K_{S0}$, $K_{S0}^{'}$). Given inconsistencies in the thermal pressure expression in \citet{anderson1994equation}, Burnman recalculates the temperature along the isentrope, the thermal pressure away from the isentrope, and the gr{\"u}neisen parameter directly from the EOS with the following thermodynamic relationships and \citet{POIRIER_book}: $\partial E/\partial S\big|_{V} = T$, $-\partial E /\partial V\big|_{S} = P$, and $\partial (\ln T)/\partial (\ln \rho)\big|_{S} = \gamma$\footnote{See e.g., \href{https://burnman.readthedocs.io/en/latest/eos.html\#anderson-and-ahrens-1994}{https://burnman.readthedocs.io/en/latest/eos.html\#anderson-and-ahrens-1994}}.

The EOS+mineral suites discussed thus far make a priori assumptions about the stable mineral phases for a given P-T either through fixed transition pressures (S07, Z16, and ZS13) or $P-T$ functions (V07, So07, Mag). ExoPlex, however, uses the \texttt{Perple\_X} software \citep{connolly_2009} to solve for the stable mantle mineral assemblage within each layer via Gibbs free energy minimization at a given P and T for input thermodynamic data and bulk composition. ExoPlex uses thermodynamic database of \citet{Stixrude_2011} which assumes thermal BM3 EOS to describe the mantle minerals, and includes pre-computed mantle grids over a wide range of mantle compositions. We adopt the default mantle grid (Si/Mg = 0.9, Ca/Mg = 0.07, Al/Mg = 0.09, and $w_\mathrm{FeO} = 0$).

\subsection{Dorn+ 2015 (D15) and Dorn+ 2017 (D17)}
\label{subsection:dorn15_and_dorn17}

\citet[][or D15]{Dorn15} uses the isothermal BM3 + thermal Mie-Grüneisen thermal EOS for $\epsilon$-Fe EOS from \citet{belonoshko_2010} whose parameters are derived from theoretical (molecular dynamic) simulations up to 1 TPa. \citet[][or D17]{Dorn_2017} adopts the Holzapfel EOS for $\epsilon$-Fe from \citet{Bouchet_2013}, which is based on ab initio molecular dynamics simulations up to 1.5 TPa. Like ExoPlex, both D15 and D17 use the \texttt{Perple\_X} software \citep{connolly_2009} and \citet{Stixrude_2011} thermodynamic database for the mantle. We adopt the default ExoPlex mantle grid for both (see Sect. \ref{subsection:exoplex}).

 \section{Direct comparison of core and mantle EOS}
 \label{sect:direct_EOS_comparisons}
In this section, we directly compare how material $\rho$ varies as a function of $P$ for the various isothermal/isentropic EOS underlying the selected EOS+mineral suites, shown in Fig. \ref{fig:P_rho_plots}. We do not incorporate the thermal EOS in this analysis for the most direct look at how pressure influences the materials' densities. We break our discussion into the EOS for core materials (Sect. \ref{sect:direct_core_EOS_comparison}), lower mantle minerals (Sect. \ref{sect:direct_LM_EOS_comparison}), and upper mantle minerals (Sect. \ref{sect:direct_UM_EOS_comparison}). In Sect. \ref{sect:effects of EOS+mineral suite on density}, we discuss how the differences in core and mantle EOS propagate to the planet level and effect the calculated $\rho_p$ of rocky planets. 

We note it is common practice in the literature to not include the uncertainties in EOS parameters when modeling the interiors of small planets. In this work, we follow suit in excluding EOS parameter uncertainty in our analyses in order to compare the EOS+mineral suites as they are presented and typically used. In order to provide a sense of scale, we point to the analysis of \citet{unterborn_panero19}, which propagates the uncertainties in EOS parameters underlying their rocky planet models over a wide range of $M_p$. The authors find the total uncertainty in the $R_p$ of a 4$M_\oplus$ planet to be $\sim0.25\%$, corresponding to 0.75\% uncertainty in $\rho_p$. Therefore, while the exact values will change between suites, the uncertainty in calculated planet density should be of order $\lesssim1\%$.

\subsection{Core EOS}
\label{sect:direct_core_EOS_comparison}
\begin{figure}[h]
     \centering
     \includegraphics[width=0.99\linewidth]{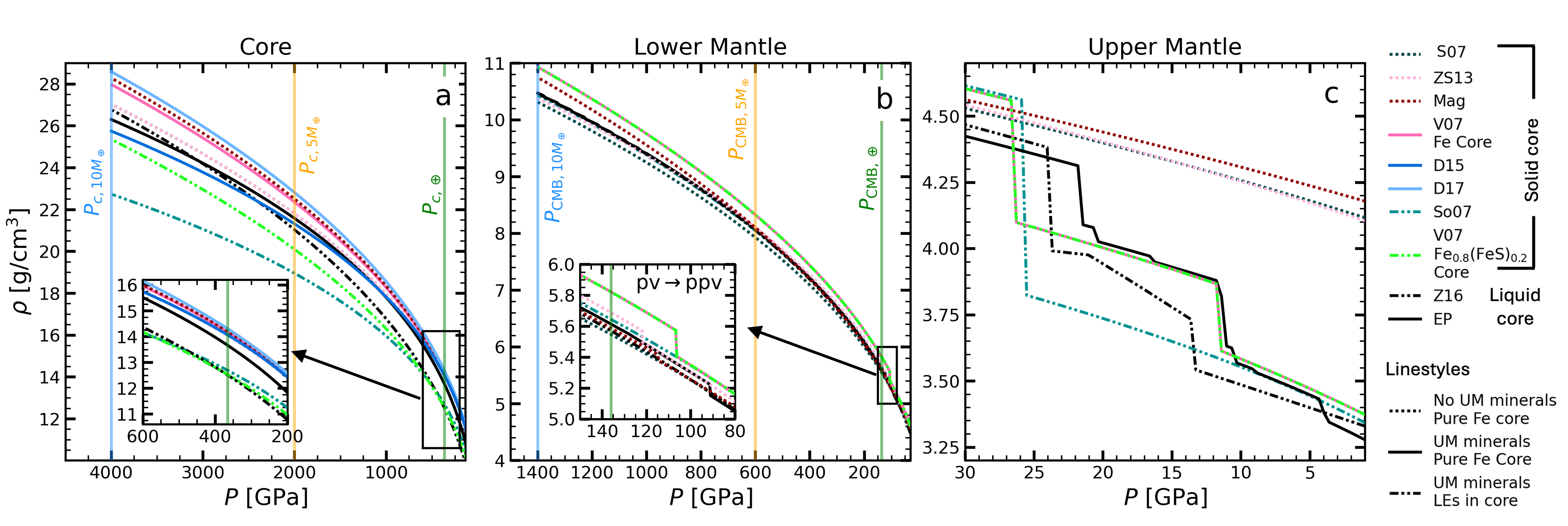}
     \caption{Material density as a function of pressure for the isothermal/isentropic (a) core and (b)-(c) mantle EOS used within the selected suites. Panel (a) spans from the central pressure of a 10$M_\oplus$ planet \citep[e.g.,][]{Boujibar_2019, unterborn_panero19} to $200$ GPa. Panel (b) shows the lower-mantle minerals from the core-mantle-boundary pressure of a 10$M_\oplus$ planet to 80 GPa. The inset shows the transition of pv$\to$ppv. Panel (c) shows the lower mantle mineral phases from 30-1 GPa. Solid lines show EOS+mineral suites that include upper mantle minerals and assume the planet core is pure Fe with no light elements (or LE, in the key). Similarly, the dashed lines show those that do not include upper mantle minerals and assume a pure Fe core with no light elements. The dashed-dotted lines show suites that include upper mantle minerals and no light elements in the pure Fe core. MAGRATHEA v1.0 is shortened to `Mag' and ExoPlex to `EP.'}
     \label{fig:P_rho_plots}
 \end{figure}

In Fig. \ref{fig:P_rho_plots}a, we show calculated material density as a function of pressure, $\rho(P)$, for the core materials adopted in the EOS+mineral suites up to the approximate central pressure ($P_c$) of a 10$M_\oplus$ Earth-like planet, or $P_{c,10M_\oplus} \sim4000$ GPa \citep[given as a vertical blue line; per][]{unterborn_panero19}. The vertical orange and green lines show the $P_{c,5M_\oplus} = 2000$ GPa and $P_{c,\oplus} = 365$ GPa values, respectively, from Sect. \ref{sect:common_eos} and Fig. \ref{fig:bm3_Vinet_comparison_scaled}. From $P\sim600-200$ GPa (Fig. \ref{fig:P_rho_plots}a inset), the EOS+mineral suites that assume the core is pure $\epsilon-$Fe (e.g., S07, ZS13, MAGRATHEA v1.0, V07 Fe, D15, and D17) give $\rho$ values that are $\sim12-13\%$ larger than those that assume some degree of light element enrichment (So07, V07 Fe$_{0.8}$FeS$_{0.2}$, and Z16). This is largely due to the $\rho_0$ values of $\epsilon-$Fe being $\sim17\%$ larger than the adopted core mixtures of So07, V07 Fe$_{0.8}$FeS$_{0.2}$, and Z16 (Table \ref{sect:selected_EOS_mineral_suites}). The 17\% difference in $\rho_0$ is somewhat larger than the 12\% in $\rho$ because $K_0$ is generally smaller for the Fe+light element mixtures ($K_0 = 135-150.2$ GPa) than for $\epsilon-$Fe ($156.2-253.8$ GPa). Put differently, the Fe+light element mixtures are more compressible than $\epsilon-$Fe which causes their $\rho$ to increase more rapidly with increasing $P$. It is for the same reasons that, despite having a lower $\rho_0$, the liquid-Fe EOS used by ExoPlex gives $\rho$ values larger than those of So07 and V07 Fe$_{0.8}$FeS$_{0.2}$ because its $K_0$ $\sim19-27\%$ is smaller. The differences between EOS for the same material (e.g., $\epsilon-$Fe) are small over the $P\sim200-600$ GPa range as it requires no/minimal extrapolation beyond the $P$ ranges used to derive their parameters (Sect. \ref{sect:common_eos} and \ref{sect:selected_EOS_mineral_suites}).

As $P\to P_{c,10M_\oplus}$, the adopted core EOS must be extrapolated well beyond the range of data used to develop each material model. As such, differences in $\rho$ between EOS for core materials become a complicated function of their respective EOS formulations (i.e., Vinet, BM3, BM2), $K_0$, and $K_0^{'}$. With the exception of Z16, the EOS for Fe+light element mixtures (So07 and V07 Fe$_{0.8}$FeS$_{0.2}$) continue to predict the lowest $\rho$ for all $P$ in Fig. \ref{fig:P_rho_plots}a.
However, despite assuming a higher $w_\mathrm{LE}$, V07 Fe$_{0.8}$FeS$_{0.2}$ gives noticeably higher $\rho$ than So07 and approaches D15, ExoPlex, ZS13, and S07 as $P\to P_{c,10M_\oplus}$. This is a combination of the fact that the V07 Fe$_{0.8}$FeS$_{0.2}$ core mixture has a lower $K_0^{'}$ and is described via a Vinet EOS, whereas So07 uses a BM3 (Fig. \ref{fig:bm3_Vinet_comparison_scaled}). Z16 notably surpasses D15 and ExoPlex around 2000-2500 GPa, and predicts nearly identical densities to ZS13/S07 at $\sim P_{c,10M_\oplus}$, due to Z16 using a BM2 EOS for the core mixture. BM2 EOS do not include a $K_{0}^{'}$ term (Sect. \ref{sect:BM3}) and therefore implicitly assume that the material's compressibility does not change for all $P$. While this assumption is fine for low pressures, it is nonphysical for the pressure scales associated with most known rocky exoplanets. Assuming characteristic $K_0^{'} = 4.5-6$ vales, the difference in $\rho$ between the Z16 core EOS and a BM3 with the same $\rho_0$ and $K_0$ is $4-10\%$ at 365 GPa and $12-27\%$ at 4000 GPa. Finally, the differences in $\rho$ for a given material are no longer small with the $\epsilon-$Fe EOS varying by $\sim10-11\%$ as $P\to P_{c,10M_\oplus}$ which is largely a result of extrapolation of the EOS to much larger pressures than those used to determine their parameters (e.g., Sect. \ref{sect:common_eos}).

\subsection{Lower Mantle}
\label{sect:direct_LM_EOS_comparison}

In Fig. \ref{fig:P_rho_plots}b, we show $\rho(P)$ from the core-mantle-boundary pressure of a 10$M_\oplus$ planet, $P_{\mathrm{CMB}, 10M_\oplus}$, to $\sim30$ GPa, the approximate transition pressure of silicate-perovskite (Fig. \ref{fig:mantle_section}). The inset highlights the silicate-perovskite to post-perovskite phase transition (labeled as pv$\to$ppv in Fig. \ref{fig:P_rho_plots}b), which is included in V07, ZS13, D15, D17, MAGRATHEA v1.0, and ExoPlex and occurs between $\sim120-95$ GPa. For pressures smaller than these transition pressures, the differences in $\rho$ are small between the suites, varying by only $\sim2\%$ and therefore is likely driven by EOS parameter uncertainties. At the core-mantle-boundary pressure of Earth, $P_\mathrm{CMB, \oplus}$ (green vertical line), this remains true with the exception of the V07 suites which predict a material density that is $\sim4\%$ larger than the rest because the post-perovskite phase is assumed to incorporate much more Fe than the other suites (Sect. \ref{sect:selected_EOS_mineral_suites}). As the pressure tends towards $P_{\mathrm{CMB}, 10M_\oplus}$, the suites that include a post-perovskite phase generally predict a higher density than those that assume silicate-perovskite is the highest pressure phase given the former mineral usually has a marginally higher $\rho_0$ than the latter (Sect. \ref{fig:mantle_section}). Like the core EOS (Sect. \ref{sect:direct_core_EOS_comparison}), however, reaching $P_{\mathrm{CMB}, 10M_\oplus}$ requires extrapolation of the material EOS well beyond the experimental/theoretical the ranges used to derive their parameters (Sect. \ref{sect:selected_EOS_mineral_suites}), leading to differences in $\rho$ of $\sim6\%$ that cannot be explained solely by the differences in $\rho_0$ between silicate-perovskite and post-perovskite.

\subsection{Upper Mantle}
\label{sect:direct_UM_EOS_comparison}

The boundary between the upper and lower mantle is defined by the silicate-perovskite transition pressure, with the upper mantle being comprised of all lower pressure mineral phases, which varies from $\sim 22-26$ GPa between the selected suites that include upper mantle minerals (So07, Z16, V07, D15, D17, and ExoPlex). To encompass the entire upper mantle regions of the selected suites, we show $\rho(P)$ from 30 -- 1 GPa in Fig. \ref{fig:P_rho_plots}c. Below 21 GPa, those suites that include upper mantle minerals naturally predict the lowest densities and, despite numerous differences in the complexity/number of minerals included, vary by only $\sim4\%$. Those that do not include any upper mantle minerals and assume bridgmanite is the lowest pressure mineral (S07, ZS13, and MAGRATHEA v1.0) agree with one another to within $\sim1\%$ but predict $\rho$ values that are $\sim15-21\%$ larger than the suites with upper mantle minerals. As such, the largest effect on average $\rho\lesssim30$ GPa is whether \textit{any} upper mantle minerals are included within the suite; whereas, differences between the selected upper mantle minerals plays a secondary role.

\section{Effects of EOS+mineral suite on rocky planet bulk density}
\label{sect:effects of EOS+mineral suite on density}

In this section, we investigate how the choice of EOS+mineral suite effects predicted rocky planet bulk density ($\rho_p$) as a function of planet mass ($M_p$) and Fe/Mg. All suites are applied using the planet builder described in Sect. \ref{sect:planet_builder}. We start by looking at differences in the material density ($\rho$) profiles of 1$M_\oplus$ and 10$M_\oplus$ planets with fixed Earth-like Fe/Mg in Sect. \ref{sect:rad_dens_profs}. We then expand the scope to variable Fe/Mg in Sect. \ref{sect:variable_femg} to explore how $\rho_p$ changes from very Fe-poor to very Fe-rich compositions. Later in Sect. \ref{sect:planet_interpretation_results}, we will show how the results of \ref{sect:rad_dens_profs}-\ref{sect:variable_femg} map onto the range of likely rocky planet interiors and influence the interpretations of small planets with precisely determined $\rho_p$ and their demographics.

\subsection{Planets With Different Masses And Moderate Earth-like Fe/Mg}
\label{sect:rad_dens_profs}

Fig. \ref{fig:profile_plots}a and b show $\rho$ as a function of radial distance from the center of 1$M_\oplus$ and 10$M_\oplus$ planets, respectively. We highlight the approximate core and mantle regions at the top of each subplot. The insets show the approximate upper mantle regions (Sect. \ref{fig:mantle_section}) of the planets. Dashed lines show the EOS+mineral suites that assume a pure-Fe core and no upper mantle mineral phases (S07, ZS13, and MAGRATHEA v1.0). Solid lines show those suites with a pure Fe core and upper mantle mineral phases: (V07 Fe, D15, D17, and ExoPlex). The suites that include light elements in the core and upper mantle mineral phases are shown via dashed-dotted lines (So07, V07 Fe$_{0.8}$FeS$_{0.2}$, and Z16). 

Our subsequent discussion focuses on material and planet bulk densities, but here we briefly discuss differences in calculated interior temperatures, namely at the CMB. With the exception of suites that are isothermal (S07 and ZS13) or otherwise do not explicitly include thermal EOS for the planet materials (Z16), the calculated CMB temperatures for 1$M_\oplus$ agree to within $\approx 200$ K, ranging from 2394-2599 K. For 10$M_\oplus$ EOS+mineral suites that include a thermal component, the differences in CMB are larger ranging from 3333-5300 K with the So07 and V07 EOS+mineral suites giving the lowest values. We note that the models shown in \citet{Sotin_2007} and \citet{Valencia_2007} include temperature discontinuities, or jumps, at material boundary layers (e.g., the CMB) whereas our planet builder does not. As such, the interior temperatures we calculate for the So07 and V07 EOS+mineral suites are lower than those presented in \citet{Sotin_2007} and \citet{Valencia_2007}, respectively. However, this will not affect the following discussion as including, for example, an 800 K temperature increase at the CMB and a 300 K increase at the upper-lower mantle boundary as done in \citet{Sotin_2007} changes the density of a 1$M_\oplus$ and 10$M_\oplus$ planet with an Earth-like Fe/Mg by $\lesssim0.5\%$.

\subsubsection{Small 1$M_\oplus$ Planets With Earth-like Fe/Mg}
\begin{figure}[h]
    \centering
    \includegraphics[width=0.99\linewidth]{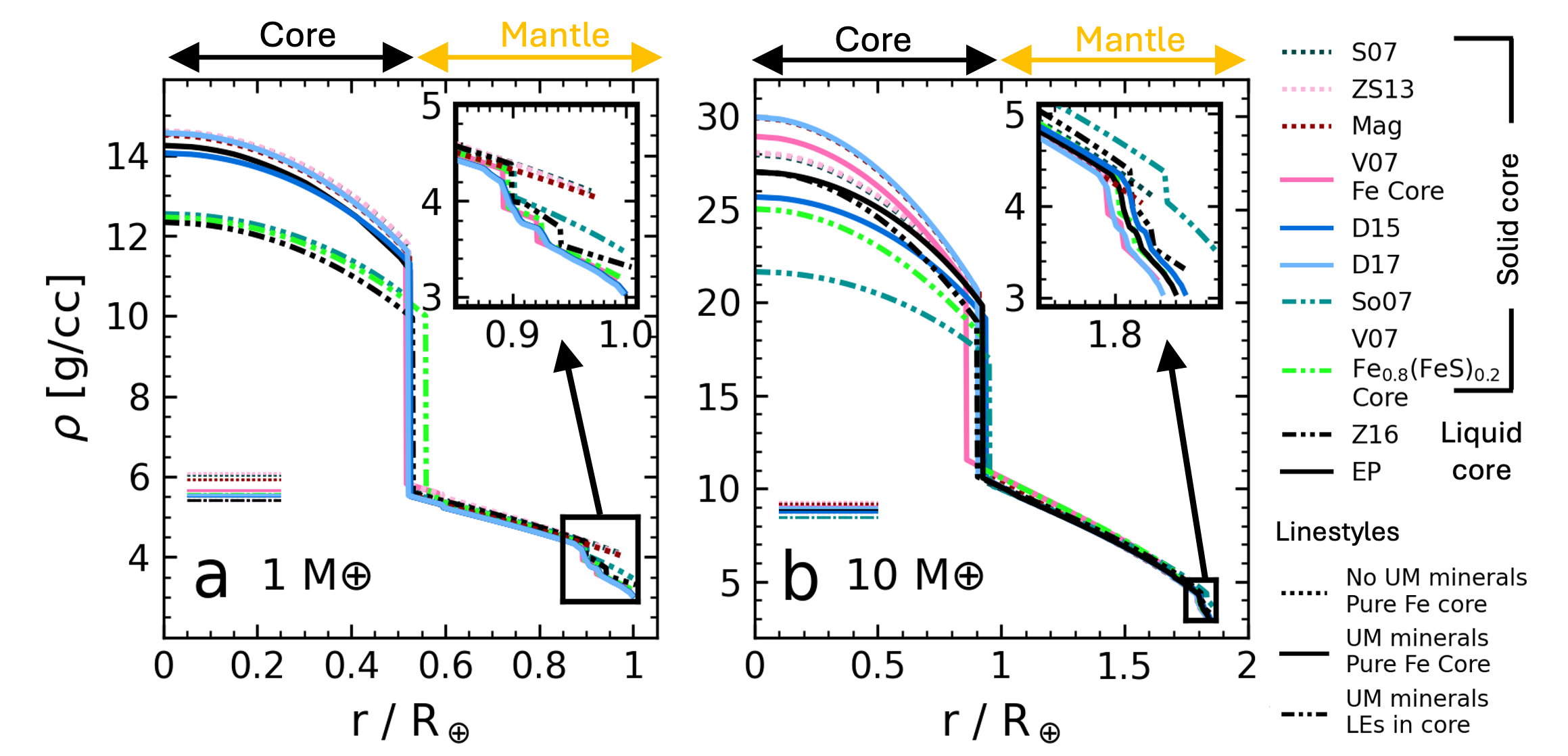}
    \caption{Material density ($\rho$) as a function of radial distance ($r$) from the center of (a) 1$M_\oplus$ and (b) 10$M_\oplus$ rocky planets with Earth-like Fe/Mg = 0.9. The various curves correspond to the EOS+mineral suites in Sect. \ref{sect:selected_EOS_mineral_suites} applied using the planet builder described in Sect. \ref{sect:planet_builder}. Suites that assume a pure-Fe core and no upper mantle mineral phases are shown via dashed lines. Solid lines show those with a pure Fe core and upper mantle mineral phases. Suites that include light elements (or LE, in the key) in the core and upper mantle mineral phases are shown as dashed-dotted lines. The horizontal lines on the left-hand side of each subplot indicate the calculated bulk density of the planets for each suite.}
    \label{fig:profile_plots}
\end{figure}

The cores of 1$M_\oplus$ planets are visually separated into two apparent categories in Fig. \ref{fig:profile_plots}a: those with light elements and those without. The EOS+mineral suites that include core light elements (V07 Fe$_{0.8}$FeS$_{0.2}$, So07, and Z16) predict average core densities that are $\sim 10-16\%$ lower than those that assume a pure-Fe core. In contrast, the suites that assume a pure $\epsilon-$Fe core (S07, V07 $\epsilon-\mathrm{Fe}$, ZS13, D15, D17, and MAGRATHEA v1.0) predict central core densities that vary by $\lesssim$ 5\% from one another. While not negligible, this difference between suites for a given core material are $\sim2-3\times$ smaller than the effects of the core composition, at least at the 1$M_\oplus$ scale. This is largely because the core pressures within a 1$M_\oplus$ Earth-like planet are generally similar to or smaller than the pressure ranges used to derive EOS parameters for core materials (Sect. \ref{sect:selected_EOS_mineral_suites}). In other words, the core EOS used in most of the selected suites do not require significant extrapolation outside their range of validity.

In order of importance, the average mantle density of a 1$M_\oplus$ planet is dictated by whether or not the EOS+mineral suite includes upper mantle mineral phases (Fig. \ref{fig:profile_plots}a inset), then the inclusion of Fe and post-perovskite. S07 and ZS13 give the highest average mantle densities of $\sim4835$ kg/m$^3$, both of which assume a Fe-free mantle \textit{without} upper mantle mineral phases. On the flip side, the suites that adopt a Fe-free mantle \textit{with} upper mantle minerals (D15, D17, ExoPlex) yield the lowest average mantle densities of 4461 kg/m$^3$, or $\sim 8\%$ smaller than S07 and ZS13. Of the suites with upper mantle minerals, those that include Fe within the mantle (So07, Z16, V07 Fe, and V07 Fe$_{0.8}$FeS$_{0.2}$) give intermediate average mantle densities that are $\sim5\%$ larger than those with no mantle Fe (i.e., $w_\mathrm{FeO}=0$). The role of post-perovskite for a 1$M_\oplus$ planet is largely negligible as expected (Fig. \ref{fig:mantle_section}a). For example, ZS13 includes post-perovskite where S07 does not, but the two otherwise make the same assumptions about the minerals present in the mantle. Despite including a post-perovskite phase, ZS13 agrees with S07 by $< 1\%$. 

While the differences between the EOS+mineral suites are a complicated function, the largest controls on $R_p$ and $\rho_p$ of a 1$M_\oplus$ Earth-like planet are the inclusion/exclusion of core light elements and upper mantle minerals. ZS13 and S07 (which have pure $\epsilon-$Fe core and no upper mantle minerals) give the smallest $R_p$ and therefore highest $\rho_p$. Z16 and So07 (which have core light elements and upper mantle minerals) give the largest $R_p$ and smallest $\rho_p$. The calculated $R_p$ values range from 0.967-1.005 $R_\oplus$ as shown in Fig. \ref{fig:profile_plots}a, while the corresponding $\rho_p$ values range from 5430-6094 kg/m$^3$ (horizontal lines), or $-0.5\%$ to $+10.5\%$ that of Earth. Our results are broadly consistent with \citet[][]{CTU16_ApJ_Scaling_the_Earth}, which investigated how perturbations in EOS parameters of different planetary materials affect the interior density profiles and masses of 1$R_\oplus$ planets.

\subsubsection{Larger 10$M_\oplus$ Planets With Earth-like Fe/Mg}

The considerations for a 1$M_\oplus$ planet remain important for the 10$M_\oplus$ planets shown in Fig. \ref{fig:profile_plots}b such as including/excluding core light elements or upper mantle minerals. However, the picture is further complicated by the extrapolation of the material EOS within the EOS+mineral suites to pressures well beyond the experimental/computation limits used to derive them (Sect. \ref{sect:common_eos} and \ref{sect:selected_EOS_mineral_suites}). This is most apparent for the core materials where there no longer exists a clear visual dichotomy between EOS+mineral suites that include/exclude light elements within the core. While So07 still predicts the lowest average core density, V07 Fe$_{0.8}$Fe$_{0.2}$ predicts an average core density similar to D15 ($\epsilon-$Fe core) while Z16 exceeds it. Even for a given material, the differences in extrapolation of EOS are quite large at the 10$M_\oplus$ mass scale. For example, the suites that adopt a pure $\epsilon-$Fe core give average core densities that vary by $\sim11\%$ in Fig. \ref{fig:profile_plots}b, more than double that of the 1$M_\oplus$ planets (Fig. \ref{fig:profile_plots}a). 

Including/excluding post-perovskite within the EOS+mineral suite becomes a non-negligible decision for 10$M_\oplus$ planets. While the jump in material density associated with the silicate-perovskite$\to$post-perovskite phase transition is small, the post-perovskite phase is expected to represent a large volume fraction of 10$M_\oplus$ (Fig. \ref{fig:mantle_section}b) which can have an appreciable impact on its average mantle density. Indeed, V07 Fe, V07 Fe$_{0.8}$FeS$_{0.2}$, MAGRATHEA v1.0, and ZS13 -- all of which include post-perovskite -- give the highest average mantle densities in Fig. \ref{fig:profile_plots}b. Of these four, both V07 suites give the highest average mantle densities as they assume the post-perovskite phase includes 20 mol \% Fe (Sect. \ref{sect:V07}) which increases its material density relative to the Mg-endmember assumed in the MAGRATHEA v1.0 and ZS13 suites.

\subsection{Rocky Planet Bulk Density as a Function of Fe/Mg}

\label{sect:variable_femg}
In Sect. \ref{sect:rad_dens_profs}, we give a detailed look at how the material density as a function of radial distance from the planet's center ($\rho(r)$) and the bulk density of planets ($\rho_p$) with moderate Earth-like Fe/Mg = 0.9 vary between EOS+mineral suites. Now we expand our analysis to a wider range\footnote{The predicted densities are calculated every $\Delta\log_{10}(\mathrm{Fe/Mg) = 0.1}$, such that the array of Fe/Mg values is $\log_{10}(\mathrm{Fe/Mg)} = -1 \;+ \;0.1\times n$ where $n\in\{0, 1, \dots20\}$.} of bulk Fe/Mg = 0.1-10. This selected Fe/Mg range spans from nearly Fe-free/core-less planets to nearly silicate-free/pure-core planets (CMF$\sim0.05-0.85$; Sect. \ref{sect:planet_builder}). Fig. \ref{fig:rho_vs_femg} shows $\rho_p$ as a function of Fe/Mg for each selected suite. Fig. \ref{fig:rho_vs_femg} (b) and (c) show the percent differences in planet bulk density, $\Delta\rho_p$, for 1$M_\oplus$ and 10$M_\oplus$ planets, respectively, defined relative to the ExoPlex suite given our planet builder is based on it. The vertical black dashed lines show the Fe/Mg ranges of the RZ (Sect.\ref{sect:planet_interpretations_and_inferences}). The green vertical line shows the moderate Fe/Mg = 0.9 adopted in Sect. \ref{sect:rad_dens_profs}. 

\subsubsection{Rocky Planet Bulk Density With Low Fe/Mg}
\label{subsect:low_FeMg}
\begin{figure}[h]
    \centering
    \includegraphics[width=0.99\linewidth]{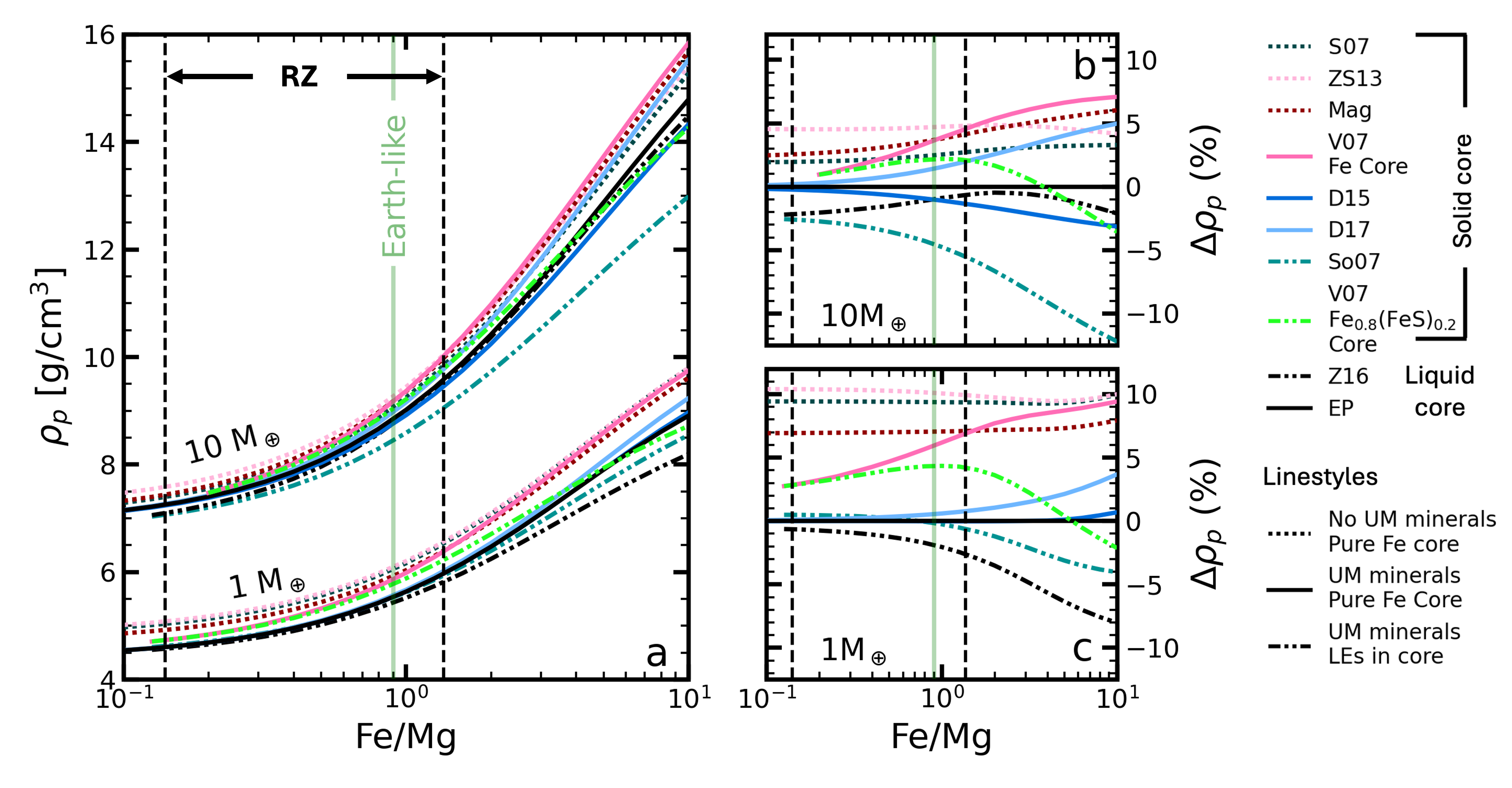}
    \caption{(a) Bulk planet density ($\rho_p$) for 1$M_\oplus$ and 10$M_\oplus$ rocky planets as a function of bulk Fe/Mg and EOS+mineral suite. Subplots (b) and (c) show the percent differences in the calculated $\rho_p$ ($\Delta\rho_p$) of each suite relative to ExoPlex at 10$M_\oplus$ and 1$M_\oplus$, respectively. Vertical lines show the Fe/Mg ranges of the RZ (black dashed; Sect. \ref{sect:planet_interpretations_and_inferences}) and Earth-like Fe/Mg = 0.9 (green; Sect. \ref{fig:profile_plots}). Suites that assume a pure-Fe core and no upper mantle mineral phases are shown via dashed lines. Solid lines show those with a pure Fe core and upper mantle mineral phases. Suites that include light elements in the core and upper mantle mineral phases are shown as dashed-dotted lines.}
    \label{fig:rho_vs_femg}
\end{figure}

In the low Fe/Mg regime, or Fe/Mg $\sim0.1$, choices about the mantle within the EOS+mineral suite play the largest role in determining planet $\rho_p$. For 1$M_\oplus$, the suites without upper mantle minerals predict $\rho_p$ values that are $\sim7-10\%$ larger than that of ExoPlex (Fig. \ref{fig:rho_vs_femg}c), while the rest agree to within $\sim3\%$. The ExoPlex, D15, and D17 suites are indistinguishable from one another as Fe/Mg$\to0$. This is because all three use the same thermodynamic equilibrium software and database to describe the mantle minerals; we also assigned them the same mantle Si/Mg, Ca/Mg, Al/Mg and $w_\mathrm{FeO}$ values (Sect. \ref{subsection:exoplex}-\ref{subsection:dorn15_and_dorn17}). The EOS+mineral suites that adopt a non-zero mantle $w_\mathrm{FeO}$ (So07, Z16, V07 Fe, and V07 Fe$_{0.8}$FeS$_{0.2}$) do not have valid solutions for 1$M_\oplus$ planets below Fe/Mg $\lesssim0.12$ as there is insufficient Fe available to satisfy their assumed mantle compositions.

For 10$M_\oplus$ planets with low Fe/Mg, the S07, ZS13, and MAGRATHEA v1.0 EOS+mineral suites still give the largest $\rho_p$ values, but are now within $\sim2-5\%$ that of the ExoPlex suite as shown in Fig. \ref{fig:rho_vs_femg}b. Meaning, the importance of including/excluding upper mantle minerals remains, but is reduced relative to the 1$M_\oplus$ case. Simultaneously, the importance of post-perovskite within the EOS+mineral suite increases with increasing mass (Fig. \ref{fig:mantle_section}). For example, So07 and Z16 -- neither of which include post-perovskite within the mantle -- give $\rho_p$ values which are $\sim2-3\%$ lower than that of the ExoPlex suite for a 10$M_\oplus$ planet as shown in Fig. \ref{fig:rho_vs_femg}b; whereas, the three differ by $<1\%$ for a 1$M_\oplus$ planet (Fig \ref{fig:rho_vs_femg}c). Like in the case of a 1$M_\oplus$ planet, the Z16 and So07 suites are ill-defined for Fe/Mg $\lesssim0.12$. However, this threshold grows to Fe/Mg $\lesssim0.2$ at 10$M_\oplus$ for V07 Fe and V07 Fe$_{0.8}$FeS$_{0.2}$ as they have a $w_\mathrm{FeO}$ value that increases with increasing planet mass (Sect. \ref{sect:V07}).

\subsubsection{Rocky Planet Bulk Density With High Fe/Mg}

In the high Fe/Mg regime, or Fe/Mg $\sim10$, choices regarding the core within the EOS+mineral suite become the dominant control on calculated planet $\rho_p$. As Fe/Mg$\to10$, the $\rho_p$ values for 1$M_\oplus$ are ordered by the phase of Fe and presence/lack of light elements within the core (Fig. \ref{fig:rho_vs_femg}a). The three suites that incorporate light elements within the core (So07, V07 Fe$_{0.8}$FeS$_{0.2}$, and Z16) give $\Delta\rho_p$ values ranging from $-2\%$ to $-8\%$ as shown in Fig. \ref{fig:rho_vs_femg}c. Of these, Z16 gives the lowest $\rho_p$ as it implicitly assumes the Fe within the core is liquid; whereas, So07 and V07 Fe$_{0.8}$FeS$_{0.2}$ assume it is $\epsilon-$Fe. On the other side, the remaining suites assume pure-Fe cores and give $\Delta\rho_p$ values ranging from 0-10\% for 1$M_\oplus$ planets. Of these, ExoPlex is the only one that assumes the pure-Fe core is in the liquid phase and it gives the lowest $\Delta\rho_p$. 

As previously seen, the picture becomes more complicated at 10$M_\oplus$ owing to the extrapolation of core EOS to pressures that greatly exceed those used to derive their parameters (Sect. \ref{fig:mantle_section} and \ref{sect:selected_EOS_mineral_suites}). In contrast to the 1$M_\oplus$ case, it is no longer \textit{strictly} true that those EOS+mineral suites with pure Fe cores give higher $\rho_p$ than those with core light elements. Most notably, Z16 does not predict the lowest $\rho_p$ for a 10$M_\oplus$ planet despite assuming a liquid-Fe core with light element enrichment. Instead, it predicts $\rho_p$ values that exceed not only So07 and V07 Fe$_{0.8}$FeS$_{0.2}$ but also D15 which assumes a pure $\epsilon-$Fe (Fig. \ref{fig:rho_vs_femg}b). This effect is largely due to the fact that Z16 adopts a BM2 EOS for the core material which does not include a $K_{0}^{'}$ (Sect. \ref{sect:BM3}). Said differently, the core EOS used in the Z16 suite neglects that materials become more resistant to self-compression with increasing pressure. As such, the core material density increases more rapidly with pressure than is physically realistic resulting in an overestimate of the core's density and $\rho_p$ as Fe/Mg$\to10$. For similar reasons, the $\Delta\rho_p$ value for So07 grows from -4\% with respect to ExoPlex for a 1$M_\oplus$ planet (Fig. \ref{fig:rho_vs_femg}c) to -12\% for a 10$M_\oplus$ planet (Fig. \ref{fig:rho_vs_femg}b). The EOS used for the Fe$_{0.87}$FeS$_{0.13}$ mixture in the So07 suite is a BM3 with a relatively large $K_0^{'} = 6$ (Table \ref{sect:selected_EOS_mineral_suites}) both of which act to noticeably reduce material density relative to Vinet EOS and/or lower $K_0^{'}$ values (Fig. \ref{fig:bm3_Vinet_comparison_scaled}).

\section{Planet Interpretations and Inferences}
\label{sect:planet_interpretations_and_inferences}
It is important to illustrate how the different EOS+mineral suites used in mass-radius-composition models can affect the interpretation, characterization, and overall demographics of small exoplanets. To accomplish this, we use the nominally rocky planet zone (RZ) classification scheme \citep{Unterborn23}. To define the RZ, we begin with the assumption that all rocky planets follow the rock-star relationship (Sect. \ref{sect:intro}) and there are no processes (such as mantle stripping) that cause its Fe/Mg to deviate from that of its host. Therefore, any diversity in the composition of rocky exoplanets is assumed to result from variation in the major rock-building elements (Fe, Mg, Si) between stars.

\begin{figure}
    \centering
    \includegraphics[width=0.8\linewidth]{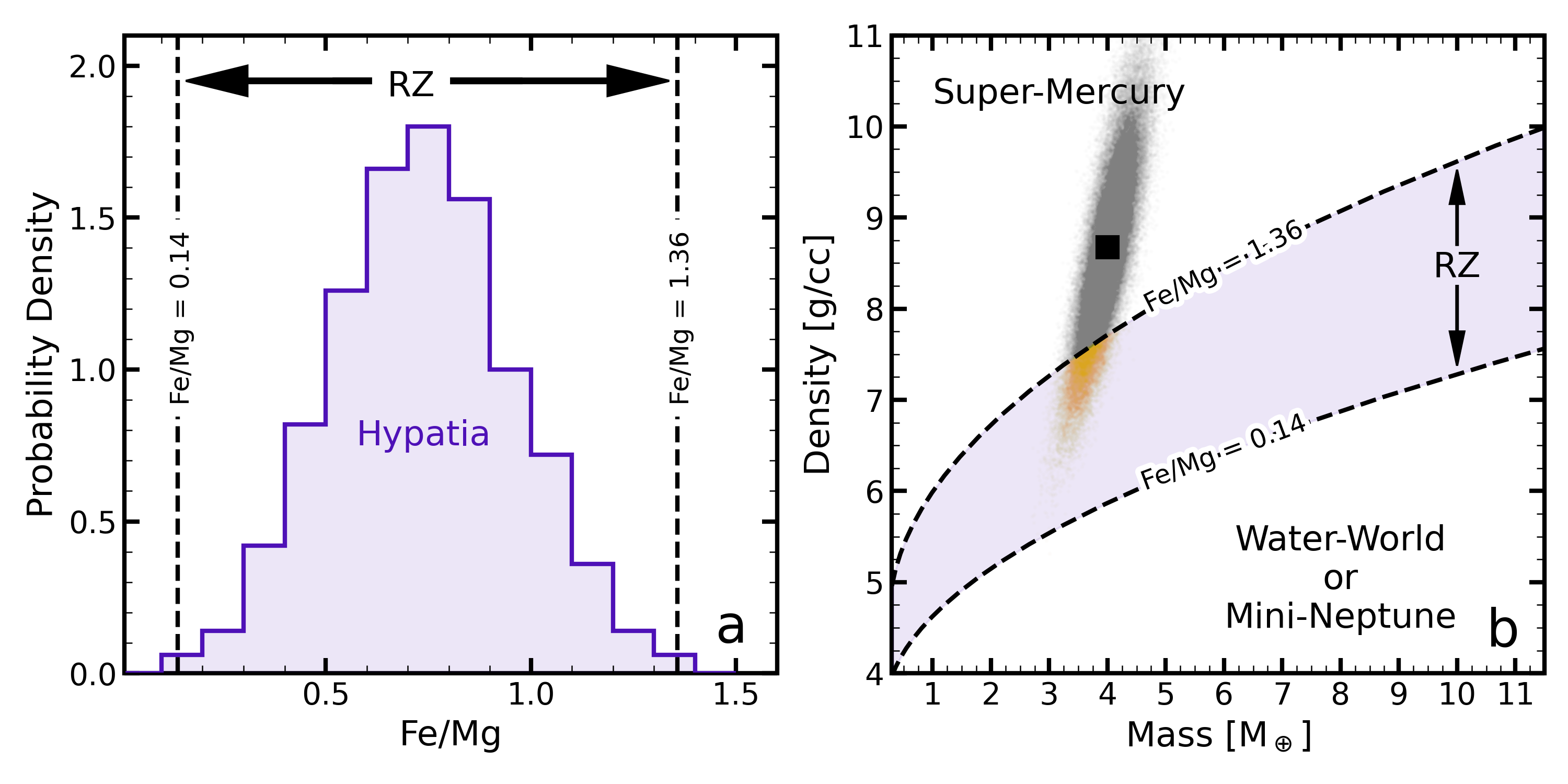}
    \caption{RZ classification scheme. Subplot (a) shows the probability density of the 500 Fe/Mg values sampled from the Hypatia Catalog (purple histogram) and the corresponding 3$\sigma$ data range (black dashed lines) which defines the compositional bounds of the RZ. Subplot (b) shows the RZ in density-mass space built using the ExoPlex EOS+mineral suite. Planets that can be statistically excluded from the RZ and have densities greater or lower than it are interpreted as super-Mercuries or water-worlds/mini-Neptunes, respectively. The gray and orange points show 10$^5$ randomly sampled mass-radius values for an example planet with $M_p = 4\pm0.32M_\oplus$ ($\pm8\%$) and $R_p = 1.37\pm0.03R_\oplus$ ($\pm2\%$). Those pairs that fall within the RZ account for 10\%, excluding it from the RZ. The remaining mass-radius pairs give planet densities that exceed that of the upper bound of the RZ, and the planet is interpreted as being a Fe-enriched super-Mercury. }
    \label{fig:RZ_schematic}
\end{figure}

We adopt the stellar abundance sample of \citet{Unterborn23} to allow for a direct comparison with the RZs and planet interpretations presented here. The sample contains 500 randomly sampled Fe and Mg abundances from the Hypatia Catalog\footnote{\href{https://hypatiacatalog.com/}{https://hypatiacatalog.com/}} \citep{Hinkel14_Hypatia} for which we calculate the molar Fe/Mg ratios \citep{Hinkel22}. In Fig. \ref{fig:RZ_schematic}a, we show the Fe/Mg distribution of the stellar sample (purple), and quantify the compositional range of the RZ as the 3$\sigma$ (99.7\%) data range (black dashed lines) or Fe/Mg = 0.14-1.36. We then input these Fe/Mg bounds into the planet builder for each selected EOS+mineral suite. In other words, we create an individual RZ for each selected suite, or a suite-specific RZ, for which we show an illustrative example in Fig. \ref{fig:RZ_schematic}b. We note that the $3\sigma$ Fe/Mg range of the RZ remains unchanged since it is defined from observational data irrespective of the chosen suite, but \textit{differences between the selected EOS+mineral suites will cause the RZ to map differently onto mass-density space}.

In order to classify a planet relative to a suite-specific RZ, we first generate 10$^5$ mass-radius samples from the planet's \textit{observed} mass, radius, and respective uncertainties, from which we calculate the planet's density distribution. We assume all planet mass and radius distributions are Gaussian. Where the planet's upper and lower mass or radius uncertainties are different, we use the average value. In Fig. \ref{fig:RZ_schematic}b, we show the 10$^5$ mass-density samples of an example planet with $M_p = 4.00\pm0.32M_\oplus$ ($\pm8\%$) and $R_p = 1.37\pm0.03R_\oplus$ ($\pm2\%$). Next, we determine the fraction of samples that fall within the bounds of the RZ which gives the probability that the planet is a member of the RZ ($P_\mathrm{RZ}$). The orange points for the example planet in Fig. \ref{fig:RZ_schematic}b show those that fall within the RZ while those in gray fall outside of it. Overall, 10\% of the samples fall within the RZ giving the planet a $P_\mathrm{RZ}$ = 0.1. Following \citet{Unterborn23}, we reject a planet as being a member of a given RZ if $P_\mathrm{RZ}< 0.3$, meaning our example planet is rejected as being within the RZ with relatively high confidence. Planets with $P_\mathrm{RZ}> 0.3$ are consistent with being a member of the RZ and the higher the $P_\mathrm{RZ}$, the more likely this is to be true. These planets are interpreted as being rocky and Earth-like insomuch as they have compositions that likely follow the rock-star relationship, but direct comparison with the compositions of their host stars are needed to confirm or refute this interpretation \citep[][]{schulze_2021, Unterborn23}. Said differently, their compositions have likely not been altered relative to the standard picture of rocky planet formation outlined in Sect. \ref{sect:intro} through giant impacts, compositional iron-silicate sorting, formation beyond the snow-line, late-stage delivery of water, etc. When excluded planets have densities less than the RZ, they require a low-density component such as significant amounts of surface liquid/solid water or substantial atmosphere and are therefore interpreted as being water-worlds or mini-Neptunes (bottom right of Fig. \ref{fig:RZ_schematic}b). Conversely, when planets have densities that exceed the RZ at their observed masses, they are interpreted as being Fe-enriched super-Mercuries, as is the case with our example planet.

We repeat this process using each selected EOS+mineral suite with the exception of V07 Fe and Fe$_{0.8}$FeS$_{0.2}$ since they are undefined for Fe/Mg $\lesssim0.2$ (Sect. \ref{sect:variable_femg}) and therefore the lower bound of the RZ (Fe/Mg = 0.14; Sect. \ref{sect:planet_interpretations_and_inferences}) is also undefined. As such, each planet will have 8 suite-specific $P_\mathrm{RZ}$ values and interpretations that may be inconsistent with one another. In other words, the observed $M_p$ and $R_p$ of a planet may be interpreted as a strong member of the RZ using one suite, pointing towards an Earth-like formation pathway, or as a strong super-Mercury candidate by another suite, pointing towards a drastically different formation history. We categorize each planet in our selected sample according to the following. When the planet's minimum $P_\mathrm{RZ}>0.3$ its inclusion within the RZ cannot be ruled out regardless of the EOS+mineral suite used to calculate the RZ. When all $P_\mathrm{RZ}$ values are $<0.3$ and the planet's density is greater or lower than the average RZ at its mass, it is interpreted as an EOS+mineral suite independent super-Mercury or water-world/mini-Neptune, respectively. Where a planet has a minimum $P_\mathrm{RZ}<0.3$ and maximum $P_\mathrm{RZ}>0.3$ it is interpreted as a suite dependent super-Mercury or water-world/mini-Neptune if its bulk density is greater or lower than the average RZ at its mass, respectively.

Planets that are consistently identified as super-Mercuries or water-worlds/mini-Neptunes regardless of the EOS+mineral suite used to determine mass-density curves of the RZ offer the best opportunity to study the diversity of small-planet formation outcomes as (1) they have robust interpretations and (2) deviate from the standard view of planet formation. While the suite independent water-worlds/mini-Neptunes can shed light on small-planet migration, late stage water delivery, atmospheric retention, and more, they require outer volatile and/or water layers to explain their observed $\rho_p$ values which is outside the scope of this work. This is, in large part, because three layer models are highly degenerate \citep[e.g.,][]{Rogers10_FeSiO3} with respect to core+mantle+water or core+mantle+atmosphere and only follow-up observations that can rule out the presence or lack of a gaseous envelope will allow for distinguishing between water-worlds or volatile-rich mini-Neptunes.

\section{Effects of EOS+mineral suite on small planet interpretations}
\label{sect:planet_interpretation_results}

We apply the RZ classification scheme outlined in Sect. \ref{sect:planet_interpretations_and_inferences} to a sample of well-characterized small planets with bulk density uncertainties of $\sigma_{\rho_p}<25\%$. We adopt the NASA Exoplanet Archive (NEA) ``Default Parameter Set"\footnote{The boolean flag indicating whether the given set of planet parameters has been selected as default by the NEA.} and only consider those planets with both measured mass ($M_p$) and radius ($R_p$). We then select only small planets defined as having $R_p<2R_\oplus$. Planets with $R_p\geq2R_\oplus$ are very likely to be mini-Neptunes, requiring an outer volatile envelope to explain their densities \citep[e.g.,][]{Fulton_2017, Neil_2020, cloutier_2020}. We find there are $\sim160$ small planets with both $M_p$ and $R_p$ measurements from which we directly calculate $\rho_p$. The planet's $\sigma_{\rho_p}$ is then found from its average $\sigma_{M_p}$ and $\sigma_{R_p}$ values via standard propagation of uncertainties \citep[e.g.,][]{taylor2004statistical}. We find there are 83 small planets that meet our definition of well-characterized, with the 15 best-characterized having $\sigma_{\rho_p}<10\%$. Our selected sample spans a wide range of $M_p = 0.326-11.1M_\oplus$, $R_p = 0.699-1.992R_\oplus$, and $\rho_p = 2.1-13.5\;\mathrm{g/cm^3}$. We reiterate that this work is focused on EOS+mineral suites for H$_2$O- and volatile-free rocky planets with solid mantles. As outlined in Sect. \ref{sect:planet_builder}-\ref{sect:selected_EOS_mineral_suites}, the mantle potential temperature is assumed to be 1600 K with the exception of suites that are isothermal (S07 and ZS13) or otherwise do not explicitly include thermal EOS for the planet materials (Z16). As such, the RZs and interpretations in this section do not account for the equilibrium temperatures of individual planets. We discuss how melting of the silicate mantle affects the bulk densities of hot, rocky planets in Sect. \ref{sect:discussion_and_recs}.

\begin{figure}[h]
    \centering
    \includegraphics[width=0.99\linewidth]{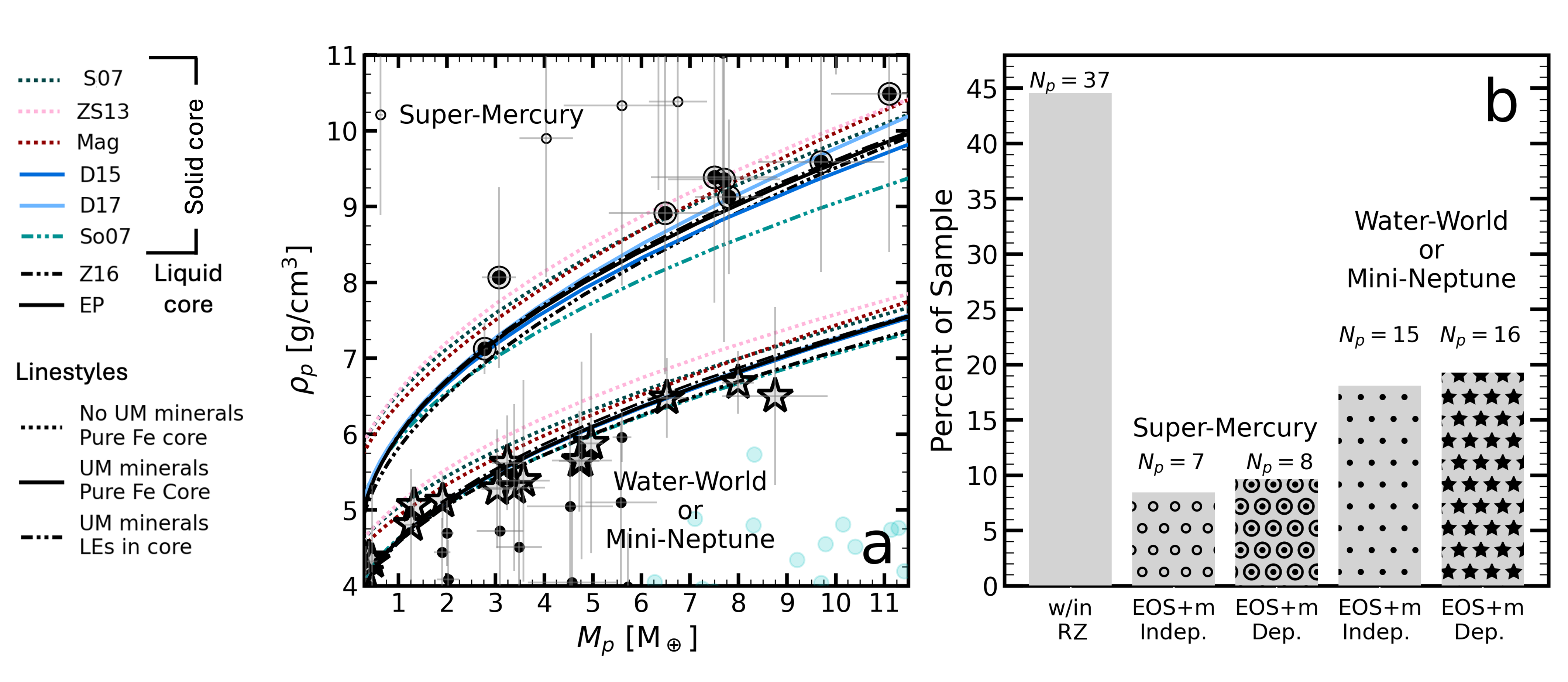}
    \caption{The RZ for each EOS+mineral suite applied using the planet builder outlined and interpretations of small planets with $\sigma_{\rho_p}<25\%$. Subplot (a) shows the calculated RZs in terms of $\rho_p$ as a function of $M_p$. Subplot (b) summarizes the planet interpretation categories outlined in Sect. \ref{sect:planet_interpretations_and_inferences}. EOS+m = EOS+mineral suite. Planets whose interpretation as a super-Mercury is independent or dependent on choice of EOS+mineral suite are shown via $\circ$ and $\odot$ symbols in both subplots, respectively. Planets whose interpretation as a water-world/mini-Neptune is independent or dependent on choice of suite are shown via $\bullet$ and $\largestar$ symbols, respectively. Planets whose inclusion within the RZ cannot be ruled out for any of the suites are labeled as ``w/in RZ'' in (b). We do not include these planets in subplot (a). The axis limits in subplot (a) are chosen to highlight the planets with suite-dependent interpretations. There are planets in the sample with suite-independent interpretations that fall outside of these bounds. The cyan points in show planets with $R\geq2R_\oplus$ which are not included in this analysis as they do not meet our definition of a small planet and are almost certainly mini-Neptunes. Suites that assume a pure-Fe core and no upper mantle mineral phases are shown via dashed lines. Solid lines show those with a pure Fe core and upper mantle mineral phases. Suites that include light elements in the core and upper mantle mineral phases are shown as dashed-dotted lines. }
    \label{fig:RZ_classifications}
\end{figure}

In Fig. \ref{fig:RZ_classifications}a, we show the RZ as calculated by the various selected EOS+mineral suites in mass-density space. Across the Fe/Mg range of the RZ, the selected suites predict values that vary by $\Delta\rho_p\pm\sim5-10\%$ (shown in Fig. \ref{fig:rho_vs_femg}b/c). These differences act to shift the RZ by amounts comparable to the $\sigma_{\rho_p}$ values of the selected planets and give contrasting interpretations for a large fraction of the sample. Fig. \ref{fig:RZ_classifications}b summarizes the number of planets ($N_p$) and corresponding percent fraction of the planet sample within each of the interpretation categories outlined in Sect. \ref{sect:planet_interpretations_and_inferences}. We find 22 planets have categorizations that are independent of the EOS+mineral suite used to calculate the RZ: 15 (18\% of sample) are suite independent water-world/mini-Neptunes ($\bullet$) and 7 (8.4\%) are super-Mercuries ($\circ$). There are 8 (9.6\% of the sample) EOS+mineral suite dependent super-Mercuries ($\odot$) and 17 (20\%) suite dependent water-worlds/mini-Neptunes ($\largestar$). This subset of EOS+mineral suite dependent planets represents $30\%$ of the total small planet sample, emphasizing the significant role the choice of suite plays when interpreting small planets at the sample level. At the individual planet level, the most drastic of these is TRAPPIST-1 d, which varies from being interpreted as a water-world with a $P_\mathrm{RZ}$ of $<7\%$ to having a $>90\%$ likelihood of being within the RZ, meaning it goes from being highly likely to be a member of the RZ to being a conclusive water-world depending on the EOS+mineral suite used.

The EOS+mineral suites that predict the highest planet densities will overestimate the number of water-worlds/mini-Neptunes and underestimate the number of Fe-enriched super-Mercuries within the selected planet sample. For example, the ZS13 suite gives the highest RZ densities over the majority of the planet sample $M_p$ (Fig. \ref{fig:RZ_classifications}a) and predicts the highest number of water-worlds/mini-Neptunes (32) and lowest number of super-Mercuries (8). Similarly, those that under-predict density will overestimate the number of super-Mercuries and underestimate the number of mini-Neptunes/water-worlds. The So07 suite generally gives the lowest $\rho_p$ values for the RZ and predicts the lowest number of water-worlds/mini-Neptunes (15) and highest number of super-Mercuries (15). There are 22 planets that fall within the two EOS+mineral suite independent categories, with 7 (7.5\% of the sample) being super-Mercuries and 15 (18\%) being water-worlds/mini-Neptunes.

\section{Discussion and Recommendations}
\label{sect:discussion_and_recs}

We review 10 EOS+mineral suites underlying rocky planet mass-radius-composition models that are common used in the literature to infer the structure and composition of small exoplanets (Sect. \ref{sect:intro}-\ref{sect:selected_EOS_mineral_suites}). The EOS used in these suites are derived from different general EOS using a variety of experimental/theoretical techniques over various pressure ranges and with varying data quality. Further, the EOS+mineral suites make many different assumptions about the materials and minerals that make up rocky planets and their phases, ranging from planets with pure solid-Fe cores and entirely bridgmanite mantles to those with light element-enriched liquid-Fe cores and Earth-like mantles. In this work, we provide the first systematic analysis of how differences between suites effect predicted rocky planet density as a function of mass and bulk composition. We apply each suite using the same planet builder (Sect. \ref{sect:planet_builder}) to solve the structure equations that govern planet interiors, thereby marginalizing over differences that may result from different computational approaches.

We find that altogether, the different choices in EOS+mineral suites lead to predicted planet bulk densities that vary by $\sim\pm10\%$ over a wide range of bulk Fe/Mg (Sect. \ref{sect:direct_EOS_comparisons}-\ref{sect:effects of EOS+mineral suite on density}). To put this into the broader context, we determine the number of confirmed small planets ($R<2R_\oplus$) listed on the NEA with both mass and radius measurements, finding $\sim160$ planets. Of these, 83 planets have density uncertainties $<$ 25\% where 15 planets have uncertainties $<$ 10\%, meaning their densities are determined with higher precision than the variability between the EOS+mineral suites underlying the mass-radius-composition models that may be used to infer their compositions. With the significant number of currently known well-measured mass+radius small planets, and more anticipated in the near future, we have entered an era where care needs to be taken when selecting a suite for modeling/interpreting small planets and direct comparison between compositional inferences of a given planet made using different suites should be done with caution. 

Here, we reiterate the differences between suites that matter the most and make recommendations for best practices when inferring/interpreting the compositions H$_2$O- and volatile-free rocky planets with solid mantles:

\begin{itemize}
\item[$\bullet$] Including/excluding core light elements has one of the largest impacts on planet density and therefore inferred rocky planet Fe/Mg (e.g., Sect. \ref{sect:variable_femg}). More than half of the selected EOS+mineral suites assume that the core is pure Fe. However, as evidenced by Earth and Mars, the two planets whose interior structures are known best, this is unlikely to be the case and there may be a large spread in the weight fraction of core light elements ($w_\mathrm{LE}$) even for planets within the same system. Therefore, at a minimum, we recommend that studies aimed at constraining the composition of a potentially rocky exoplanet do so for both $w_\mathrm{LE} = 0$ and $w_\mathrm{LE} = 20\%$, corresponding to the pure-Fe core scenario and a Mars-like light element enrichment, respectively. Following this recommendation will yield a range of Fe/Mg values for the planet that account for the (1) potential wide spread in the $w_\mathrm{LE}$ as evidenced by the Solar System and (2) the large impact that core light elements have on rocky planet density as shown in this work.

\item[$\bullet$] There is also the issue of whether the core Fe is liquid or solid with the majority of the suites opting for the latter and describing the core via the solid $\epsilon-$Fe phase. Mercury, Earth, and Mars are all believed to have a solid inner core surrounded by a liquid outer core \citep[e.g., ][]{PREM, genova2019geodetic, pommier2025experimental}. While this may also be the case for many rocky exoplanets, it is a complicated function of many planet parameters such as the identity and weight fraction of core light elements, initial temperature profile, abundance of radiogenic heating elements, mass, and age \citep[e.g.,][]{Boujibar_20, Nimmo_2020, Bonati_2021, Kraus_2022}. In short, the degree to which the cores of known, likely rocky exoplanets are liquid versus solid is unknown. Therefore, when inferring the Fe/Mg of a rocky planet, we recommend either doing calculations for both the entirely liquid and entirely solid core cases, or including a melting cure for Fe that accounts for variable amounts of core light elements like that of \citet{BICEPS}.

\item[$\bullet$] Of the various mantle considerations, including/excluding upper mantle mineral phases has the largest impact on planet bulk density. While water worlds and mini-Neptunes may not have upper mantles owing to the potentially high pressures at the H$_2$O-mantle or atmosphere-mantle boundary, all water- and volatile-free rocky worlds will. As such, we highly recommend that all EOS+mineral suites include at least the dominant upper mantle mineral phases: olivine, orthopyroxene, wadslyite, and ringwoodite. To second order, we recommend using a software like Perple\_X (e.g., Sect. \ref{subsection:exoplex}-\ref{subsection:dorn15_and_dorn17}) that incorporates a large number of possible mineral phases and determines the equilibrium assemblage for a user-specified composition to explore a range of mantle compositions given the wide spread in stellar Si/Mg and $w_\mathrm{FeO}$ between Mars and Earth. Similarly, it allows for host star abundances (where available) to be used as constraints on the mantle mineralogy and therefore planetary dynamics. Differences between suites that include post-perovskite and those that do not is negligible for 1$M_\oplus$ planets and remains relatively minor even for 10$M_\oplus$ planets with large mantles ($\Delta\rho_p\sim2-3\%$; Sect. \ref{subsect:low_FeMg} and Fig. \ref{fig:rho_vs_femg}), meaning it should not have a significant impact on the inferred Fe/Mg of most \textit{currently known} likely rocky exoplanets. Regardless, it is best practice to include the silicate-perovskite to post-perovskite phase transition as it is experimentally confirmed \citep[e.g.,][]{Murakami_2004, Oganov_2004} and the number of confirmed higher mass ($\sim10M_\oplus$), likely rocky exoplanets with precisely determined densities will only increase.

\end{itemize}

Lastly, we discuss several points that fall outside the scope of this work but are, nonetheless, important considerations for constraining/interpreting the compositions of small exoplanets:

\begin{itemize}
\item[$\bullet$] As noted in Sect. \ref{sect:intro}, the EOS+mineral suites considered in this work assume that the mantles of the rocky planets are entirely solid. This may not be the case for rocky planets with short orbital periods (up to a few days) and with, therefore, surface temperatures that likely exceed the melting temperatures of silicates, implying their mantle is at least partially liquid. \citet{Boley_2023} finds that the effects of silicate melting has a small impact on the bulk densities of atmosphere-free rocky planets with $M_p\gtrsim2-3M_\oplus$, where the density of magma ocean planets are consistent with their solid mantle counterparts to within $\sim1\%$ and $\sim2\%$ for surface temperatures of 1600 and 2000 K, respectively. However, for a 1$M_\oplus$ planet, these differences grow to $\sim3-5\%$ and $\sim8-10\%$ for 1600 and 2000 K surface temperatures, respectively. As such, we recommend future rocky planet mass-radius-composition models incorporate phase diagrams for the mantle that include silicate melts like those included in MAGRATHEA and BICEPS, particularly when attempting to constrain the compositions of short-orbital period rocky planets with $M_p\lesssim1M_\oplus$.

\item[$\bullet$] Studies that present a mass-radius-composition model or use one to infer the composition of a planet often do not account for how the uncertainties in the underlying EOS parameters affect the overall analysis. While \citet{unterborn_panero19} show that the effect of quoted EOS parameter uncertainties generally has a much smaller impact on planet density than compositional considerations, it is still best practice to propagate the uncertainties in the selected material EOS. However, a larger issue is that not all EOS for a given material are equally reliable, as differences in experimental or computational techniques can produce systematic offsets and inconsistent density estimates. All the models assemble material-dependent equations of state from either experiments or calculations. Each approach has systematic errors in their collection, and the reliability of the EOS is therefore dependent upon the quality of bias mitigation. Experimental values are generally attained through by X-ray diffraction measurements of the material under static compression. All solid samples are therefore susceptible to non-hydrostatic stresses in the sample chamber, often able to withstand several GPa in deviatoric stresses, leading to non-uniform strains, in which the lattice planes are systematically understrained sub-parallel to the principle direction of stress. The most common measurement geometry is for the X-ray to be co-linear with the principle direction of stress. In measurements for which non-hydrostatic stresses persist in the sample chamber, the X-ray then reflects off the lattice plane sub-parallel to the principle direction of stress, underestimating the volumetric strain at a given stress state. This leads to an equation of state that is artificially stiffer with a reduced pressure derivative of the bulk modulus over the pressure range of the experiment. Upon extrapolation, this leads to a significant underestimate in density \citep[][]{singh1998analysis}.

\item[$\bullet$] A key contributor to determining the makeup of an exoplanet are the stellar abundances, in lieu of any direct composition measurements (Sect. \ref{sect:intro}). However, high precision abundances are often determined for relatively small samples (100s of stars) that focus on a limited number of elements. The Hypatia Catalog is the largest database of elemental abundances for stars near to the Sun as compiled from literature \citep{Hinkel14_Hypatia}, which makes it easy to access most stellar compositional information within the field. But Hypatia is inherently a heterogeneous database, combining data from a variety of epochs, telescopes, instrument resolutions, and abundance methodologies \citep[currently containing $\sim$360 datasets,][]{Hinkel22}. The uncertainties associated with the different datasets are also calculated in non-comparable ways, such that the errors within the Hypatia Catalog are determined from the \textit{spread} or range in the available data as the truest way of determining how ``well" a value is agreed upon \citep{Hinkel14_Hypatia}. For those who are interested in minimizing associated stellar abundance errors when analyzing the interior of specific planets, we recommend using only the data that is high resolution (e.g., $>$ 100,000) and signal-to-noise ($>$ 100), where the abundance methodologies are stellar physics-driven\footnote{A PDF of dataset metadata -- including telescope and methodology details -- is provided through the Hypatia ``Help" page. In addition, specific catalogs can be viewed/plotted/downloaded using the ``Allow only catalogs" function available on any data page.}, as opposed to data-driven. In addition, since most stellar models are based on the Sun, abundances of solar twins (whose key properties only vary somewhat compared to the Sun) are typically considered the most precise abundances. 

\item[$\bullet$] When considering target planets and their host stars, we have a made a point of focusing on FGK-type stars, that have effective temperatures between 4000--6500 K. However, M-dwarf stars -- which are the smallest and most prevalent stars in the galaxy \citep{Henry18} with temperatures between 2500--4000 K -- are the best location for finding smaller, Earth-sized planets since exoplanet detection techniques are typically dependent on star-to-planet mass and radius ratios \citep{Dressing15, Hinkel24}. In other words, it is much easier to find a smaller planet around a smaller star. The most famous example of an M-dwarf planetary system is TRAPPIST-1, which boasts 7 small planets with 3 in the habitable zone \citep{Gillon17}. Unfortunately, it is extremely difficult to measure the abundances of M-dwarf stars from the ground because their stellar spectra are optimized in the infrared, which is almost entirely blocked by the Earth's atmosphere (specifically O$_2$, O$_3$, CO$_2$, and H$_2$O). In addition, because of their low temperatures, molecules as well as individual elements are observable within M-dwarf spectra, which further complicate abundance determination. As a result of these multiple issues, only $\sim$400 stars, or $\sim$3\% of the total, within the Hypatia Catalog are M-dwarfs, for which there are only 2-3 element measurements per star. This explains the wide range of classification for TRAPPIST-1 d, that was discussed in Sect. \ref{sect:planet_interpretation_results}. Unfortunately, given the dearth of M-dwarfs abundances in general, it is not currently possible to construct an M-dwarf-specific RZ (or M-RZ) in order to help in M-dwarf planet classification. It is only through a dedicated abundance survey above the atmosphere \citep[for example, the proposed Spectroscopic Abundances to Know the Heritage of M-dwarf Environs through Time -- or SAKHMET -- balloon mission:][]{Hinkel_AAS_abstract_2024} that we can hope to achieve the depth and breadth of M-dwarf abundances so that we can better characterize their planets to the same consistency and accuracy as planets hosted by FGK-type stars. 
\end{itemize}

\section{Acknowledgments}
We thank the anonymous reviewers for their comments that have improved this paper. JGS and NRH are partially supported by a NASA ROSES-2020 XRP Grant (20-XRP20 2-0125) and an ICAR-2022 Grant (22-ICAR22\_2-0016). JGS would like to thank Kameron Gausling for his useful comments which increased the clarity of this paper, as well as Lupin, Newt, Aero, and Sam for their support. NRH would like to thank CHW3, Tatertot, and Lasagna. The research shown here acknowledges use of the Hypatia Catalog Database, an online compilation of stellar abundance data as described in Hinkel et al. (2014, AJ, 148, 54). This research has made use of the NASA Exoplanet Archive, which is operated by the California Institute of Technology, under contract with the National Aeronautics and Space Administration under the Exoplanet Exploration Program. This material is based upon work supported by (while serving at) the National Science Foundation. Any opinion, findings, and conclusions or recommendations expressed in this material are those of the author(s) and do not necessarily reflect the views of the National Science Foundation or of the Federal government. The authors acknowledge support from NASA grant 80NSSC23K0267, funded through the NASA Exoplanets Research Program, PI Cayman Unterborn. The results reported herein benefitted from collaborations and/or information exchange within NASA’s Nexus for Exoplanet System Science (NExSS) research coordination network sponsored by NASA’s Science Mission Directorate, grant 80NSSC23K1356, PI Steve Desch.

\newpage
\bibliography{biblio}{}
\bibliographystyle{aasjournal}


\end{document}